\documentclass[conference,compsoc,letterpaper]{IEEEtran}
\usepackage{amsmath}
\usepackage{graphicx}
\usepackage{multirow}
\usepackage{tabularx}
\usepackage{booktabs}
\usepackage{xcolor}
\usepackage{subfig}
\usepackage{tcolorbox}
\usepackage{pifont}
\usepackage{fontawesome}
\usepackage{url}

\usepackage[all=normal,paragraphs=tight,floats=tight,leading=tight]{savetrees}

\newcommand{\emptydot}{\faCircleO}
\newcommand{\halfdot}{\faAdjust}
\newcommand{\fulldot}{\faCircle}
\newcommand{\cmark}{\ding{51}}%

\title{ Revealing the True Indicators: \\ Understanding and Improving IoC Extraction From Threat Reports

}

\author{
    \IEEEauthorblockN{
        Evangelos Froudakis\IEEEauthorrefmark{1},
        Athanasios Avgetidis\IEEEauthorrefmark{1},
        Sean Tyler Frankum\IEEEauthorrefmark{2},\\
        Roberto Perdisci\IEEEauthorrefmark{2},
        Manos Antonakakis\IEEEauthorrefmark{1},
        Angelos D. Keromytis\IEEEauthorrefmark{1}
    }
    \IEEEauthorblockA{\IEEEauthorrefmark{1}Georgia Institute of Technology\\
    \{efroudakis3, avgetidis, manos, angelos\}@gatech.edu}
    \IEEEauthorblockA{\IEEEauthorrefmark{2}University of Georgia\\
    \{Sean.Frankum, perdisci\}@uga.edu}
}

\begin{document}

\maketitle

\begin{abstract}
Indicators of Compromise (IoCs) are critical for threat detection and response, marking malicious activity across networks and systems. Yet, the effectiveness of automated IoC extraction systems is fundamentally limited by one key issue: the lack of high-quality ground truth. Current extraction tools rely either on manually extracted ground truth, which is labor-intensive and costly, or on automated ground truth creation methods that include non-malicious artifacts, leading to inflated false positive (FP) rates and unreliable threat intelligence.
In this work, we analyze the shortcomings of existing ground truth creation strategies and address them by introducing the first hybrid human-in-the-loop pipeline for IoC extraction, which combines a large language model-based classifier (LANCE) with expert analyst validation.
Our system improves precision through explainable, context-aware labeling and reduces analysts' work factor by 43\% compared to manual annotation, as demonstrated in our evaluation with six analysts. Using this approach, we produce PRISM, a high-quality, publicly available benchmark of 1,791 labeled IoCs from 50 real-world threat reports. PRISM supports both fair evaluation and training of IoC extraction methods and enables reproducible research grounded in expert-validated indicators.
\end{abstract}

\section{Introduction} \label{sec:Intro}

Detecting and analyzing malicious activity is essential in modern cybersecurity defense. Indicators of Compromise (IoCs), such as IP addresses, domain names, URLs, and file hashes, which signal malicious activity, are among the most important parts of threat intelligence \cite{GoodFATR,STIOCS,RansomIoC}. Analysts rely on IoCs for both immediate defense, such as creating block lists and firewall rules, and for activities such as threat hunting. Accurate and timely identification of IoCs can therefore significantly improve an organization's ability to detect, respond to, and mitigate cyber threats. 

One of the most prominent sources of IoCs is threat reports (TRs), which cybersecurity companies, research institutions, and government agencies publish to share critical information about attacks, campaigns, and incidents \cite{Survey1,STIOCS,Collecting_Indicators, GoodFATR}.
Threat reports vary widely in structure and detail due to the lack of standardized formatting.
This inconsistency makes accurate IoC extraction challenging, as reports may embed indicators in diverse formats and contexts, yet their precise identification is critical for effective threat response.

Previous work has focused on three general approaches for extracting IoCs from threat reports: manual labeling, automated extraction, and automated retrieval from threat intelligence sharing platforms.
As noted repeatedly~\cite{CTI_View,GoodFATR,iACE,ChainSmith}, manual labeling is the most accurate approach to extract IoCs from unstructured threat reports. However, this requires significant domain knowledge and can be very time-consuming, making it expensive and difficult to scale \cite{GoodFATR,CTI_View,iACE}. 
Automated IoC extraction tools, including recent work leveraging large language models (LLMs)~\cite{llm1, llm2, llm3}, offer potential efficiency but often lack the necessary accuracy due to the complexity of natural language in technical cybersecurity reports, the variability in IoC presentation, and the difficulty in distinguishing true IoCs from similar-looking but benign artifacts (e.g., indicators such as benign domain names, IP addresses, and URLs that are not related to the threats described in the reports). 
Finally, while there exist efforts to distribute IoCs to the community in an easy-to-parse structured format, for instance via threat exchanges (e.g., AlienVault~\cite{AlienVault}), threat report databases (e.g., \url{orkl.eu}), and online security scanners (e.g., VirusTotal~\cite{virustotal,VT_survey1,VT_survey2,VT_survey3}), we find that popular threat report IoC exchanges, like AlienVault, often automatically extract IoCs from unstructured TRs with low precision ($<80\%$), recall ($<80\%$), and/or coverage.
As a result, these approaches either incur high manual work factor costs or yield unreliable intelligence, underscoring the need for a new solution that is both accurate and efficient. 

To address these challenges, we propose the first IoC extraction approach that combines automated extraction supported by explainable AI with {\em human-in-the-loop} verification.
Our central hypothesis is that large language models (LLMs), when {\em guided by properly engineered prompts and paired with analyst feedback}, can match or surpass the accuracy of manual-only labeling while significantly reducing the time cost of human effort. In our evaluation, we observe a 43\% reduction in parsing time compared to manual-only workflows.
Importantly, our pipeline design and prompt engineering strategies are model-agnostic and generalize across most state-of-the-art LLMs, enabling flexible deployment in diverse environments.
By integrating human analysts in the IoC extraction process, we can leverage the speed of automation while ensuring high precision through human judgment. 
Similar ``human-in-the-loop'' (HITL) approaches have been successfully demonstrated in other domains where high-accuracy data labeling and expert oversight are critical, such as medical applications and some computer vision and natural language processing tasks~\cite{HITL1,HITL2,HITL3,HITL4}. To the best of our knowledge, we are the first to adapt the HITL approach for the efficient and accurate labeling of IoCs from unstructured reports. To this end, we develop a HITL system that provides: 
(1) visual assistance (via highlighted text) that allows analysts to quickly identify indicators of interest in unstructured (potentially long) threat reports; and 
(2) color-coded IoC labels assigned by an LLM, each accompanied 
by an explanation of the contextual cues used to justify the label, designed following principles of effective data visualization \cite{UI1,UI2,UI3}. 

\begin{figure}

    \resizebox{\linewidth}{!}{%
    \centering
    \includegraphics[width=\linewidth]{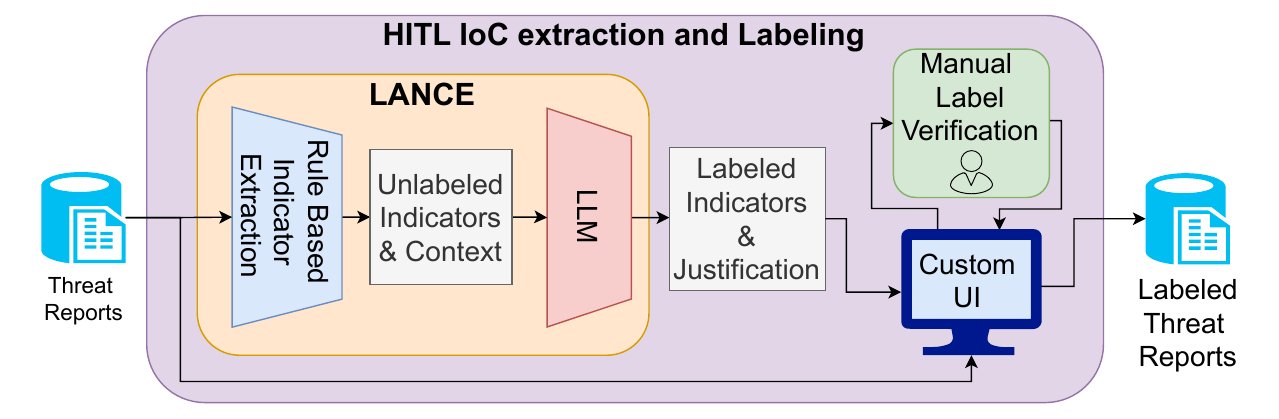}
    }
    \caption{Overview of the human-in-the-loop (HITL) IoC extraction pipeline. It combines automated extraction and LLM-based labeling (LANCE) with manual annotation, thus promoting efficient, high-confidence IoC labeling.}
    \label{fig:TotalPipeline}
\end{figure}

Figure~\ref{fig:TotalPipeline} shows an overview of our system. It consists of LANCE and a human-in-the-loop (HITL) component where analysts, supported by LANCE and by a custom web-based user interface, assign labels to IoCs found in threat reports. Thanks to our system's custom user interface and explainable LLM-based IoC labels, analysts can quickly confirm or correct the label recommendations throughout the report, thus boosting accuracy and efficiency.
To evaluate our approach, we applied it to 50 real-world threat reports from reputable sources and observed that junior analysts made fewer labeling errors when assisted by our system. We then used the system to construct PRISM, a high-quality, manually validated dataset of 1,774 labeled IoCs, including domains, URLs, IPs, and file hashes. PRISM supports both training and evaluation of IoC extraction tools and serves as the first openly available benchmark for this task. 



In summary, the main contributions of this work are:
\begin{itemize}
    
    \item \textbf{Human-in-the-loop IoC Labeling System:} The first human-in-the-loop system for IoC extraction and labeling from threat reports, which combines an LLM-based automated and explainable IoC labeling process with manual verification to minimize labeling costs and maximize accuracy.

    \item \textbf{LANCE:} The LLM–based labeling component that provides explainable classifications of indicators from unstructured threat reports. This explainability allows for faster human validation, reducing annotation time by 43\% compared to manual labeling. 
    
    \item \textbf{Review of Existing IoC Extraction/Labeling Systems:} We review and attempt to reproduce previous work on IoC extraction and labeling, including automated IoC extraction systems and threat exchanges. Our results show that existing available systems (specifically, systems whose code is available and reusable, or can be accessed via an API) either lack precision (i.e., high false positives) and/or recall (i.e., high false negatives).

    \item \textbf{PRISM:} The first open, fully manually validated, and well-documented benchmark dataset of IoCs from threat reports, consisting of 1,791 indicators extracted from 50 real-world threat reports. Our dataset, called PRISM, is designed to both future training and testing of IoC extraction/labeling models and serve as a common evaluation dataset for comparing IoC extraction systems among each other.

\end{itemize}
    
The HITL system, LANCE, and PRISM are freely available at https://github.com/EvanFr/LANCE

\section{Background and Related Work}
\label{sec:Background}

In this section, we analyze the challenges that affect automated IoC extraction and the implications of IoC misclassifications. We then present the IoC ground truth generation methods used in recent works.

\subsection{IoC Extraction Challenges and Previous Work}
\label{subsec:IoCExtractionDifficulties}

Extracting Indicators of Compromise (IoCs) from unstructured threat reports is far from straightforward. While IoCs like IP addresses, domains, URLs, and file hashes are vital for threat detection and response, they are often buried in reports that lack consistent formatting. These indicators may appear inline, in tables, or within images, and their relevant context may be scattered throughout the document~\cite{ChainSmith}. This fragmented presentation makes it difficult for automated tools to  detect and interpret IoCs reliably.

A further complication lies in differentiating true IoCs from benign artifacts that share a similar structure~\cite{AspIoC,Collecting_Indicators}. Regular expression–based extractors, though widely used, frequently overreport by flagging non-malicious entities such as popular domains embedded within malicious URLs~\cite{GoodFATR}. Because maliciousness is context-dependent, accurately identifying IoCs requires understanding context that is not always located near the indicator itself, making it difficult for NLP methods to draw accurate associations.

To address these challenges, researchers have proposed a range of extraction methods, spanning rule-based approaches~\cite{GoodFATR,STIXnet}, NLP~\cite{ChainSmith,TIMiner,Two-phase_System}, machine learning (ML)~\cite{IoCMiner,Twiti,IoCStalker,Extracting_actionable_information}, deep learning (DL)~\cite{iACE,AITI,AspIoC,STIOCS,CTI_View,Collecting_Indicators,Automatic_Identification}, including LLMs~\cite{llm1,llm2,llm3}, and graph-based techniques~\cite{HINTI,HinCTI}. However, each approach has its drawbacks. Rule-based systems lack contextual awareness. NLP models often struggle with inconsistent grammar, distant dependencies, and report structure~\cite{Twiti}. ML and DL techniques require large, accurately labeled datasets, which are scarce~\cite{ChainSmith,CTI_View,STIOCS}. LLM-based approaches are prone to hallucinations and are limited by their context window \cite{llm1, llm2, llm3, hallucinations, LLMsurvay}.

The most reliable and commonly used method remains manual labeling \cite{GoodFATR,STIXnet,ChainSmith,Two-phase_System,Twiti, IoCMiner,Extracting_actionable_information,Collecting_Indicators,iACE,STIOCS,AITI,Automatic_Identification,CTI_View,HinCTI,HINTI}. Unfortunately, despite its higher accuracy compared to automated extraction, it is expensive, time-consuming, and does not scale to meet the growing volume of threat intelligence~\cite{GoodFATR}. Similarly, Cyber Threat Intelligence (CTI) platforms, which aim to streamline indicator sharing, often provide IoCs without essential contextual metadata, such as attack stage or malware behavior, that is necessary for assessing relevance~\cite{TIMiner}. 

{In addition to IoC-focused extraction systems, recent works have broadened CTI extraction to include relations and attack behavior. For instance, Extractor~\cite{extractor}, Looking Beyond IoCs (LADDER)~\cite{LADER}, and CTINexus~\cite{CTINexus} leverage NLP and LLMs to extract complex threat intelligence beyond indicators (e.g., attack patterns and knowledge graphs). While these systems offer a broader scope, they rely heavily on RegEx for indicator extraction or do not target high-precision IoC labeling. For example, CTINexus uses subject–verb–object triplet extraction with LLMs but exhibits low recall in our replication (see Appendix~\ref{app:CTI_eval}). In contrast, our work focuses on precise, context-aware IoC labeling and complements these systems by improving the fidelity of extracted indicators.}

Our HITL system is different from previous work in that it addresses the above challenges by providing LLM-generated, context-aware IoC labels to human analysts who can quickly validate or correct ambiguous labels with full visibility into the broader document structure.

\subsection{IoC Ground Truth Generation Methods}
\label{subsec:GTgenerationMethods}

Our HITL approach integrates analyst review into the pipeline to improve IoC labeling fidelity and enable the generation of high-quality ground-truth benchmark datasets.

Incorrect ground truth (GT) data, with mislabeled benign/malicious indicators, harms both training and evaluation of automated systems~\cite{betterGT1,betterGT2}. Models trained on noisy labels risk learning incorrect patterns, overfitting, or failing to generalize. For example, labeling \texttt{github.com} as malicious teaches the model to overreport benign indicators, while missing actual threats reduces future detection capability. Ambiguity in what constitutes maliciousness, often context-dependent, further complicates matters. 

In practice, misclassifications carry real-world consequences~\cite{FP_impact1,FP_impact2,adobe}. False positives increase alert fatigue and waste analyst effort, while false negatives let threats bypass detection. These mistakes degrade the efficacy of threat detection systems and damage trust in automation. Addressing them requires both better extraction methods and access to clean, well-labeled datasets that reflect real operational challenges.

To understand how IoC ground truth has been generated so far by the research community and extract useful lessons, we conducted an extensive survey of peer-reviewed papers focused on extracting IoCs from unstructured threat reports. We selected papers that use diverse techniques for extraction and IoC extraction is one of their main contributions. Table~\ref{tab:ioc_tools} summarizes each method's technique, GT strategy, dataset size, availability of the code and dataset, and usability of any available code. For all the tools for which the code and/or the dataset were not available, we contacted the authors twice in an attempt to obtain them. Our analysis reveals a lack of transparency in GT construction and limited availability of datasets and code, hindering reproducibility and fair comparison. 

Based on our extensive literature search, we identified six common GT creation strategies: manual labeling, VirusTotal (VT)-based validation, RegEx-based matching, curated lists, unsupervised methods, and forum-based sources.

\textbf{Manual Labeling:} The vast majority of prior works rely on manual GT creation, with over half using it exclusively. Manual efforts ensure precision but are time-consuming and unscalable. Tools across all categories, including rule-based systems \cite{GoodFATR,STIXnet}, NLP-based systems \cite{ChainSmith,Two-phase_System}, machine learning (ML) systems \cite{Twiti, IoCMiner,Extracting_actionable_information}, deep learning (DL) models \cite{Collecting_Indicators,iACE,STIOCS,AITI,Automatic_Identification,CTI_View}, and graph-based approaches \cite{HinCTI,HINTI}, use manual labels to train or validate their systems. Despite the high precision, this approach lacks scalability.

\textbf{VirusTotal-Based Validation:} More than one-third of the works, mainly ML systems \cite{IoCMiner,Twiti, IoCStalker, Extracting_actionable_information} but also DL models \cite{AspIoC} and graph-based implementations \cite{HinCTI}, use VirusTotal (VT) to validate extracted IoCs. While VT offers broad coverage, threshold choice is inconsistent across studies, and VT can introduce false positives or negatives \cite{VT_survey1,VT_survey2,VT_survey3}.

\textbf{RegEx-Based Matching:} Widely used in both extraction and GT creation, regular expressions (RegEx) provide high recall but poor precision due to limited contextual awareness (e.g., IoC Searcher~\cite {GoodFATR}). Most systems use RegEx in conjunction with other methods to improve accuracy \cite{TIMiner,IoCStalker,iACE,AspIoC,STIOCS,CTI_View}.

\textbf{Threat Exchanges and other Information Sharing Platforms:} Some systems enrich their datasets with indicators gathered from forums or open-source threat sharing platforms (e.g., AlienVault)~\cite{iACE,STIOCS,CTI_View,HinCTI}. However, these forums are often noisy and unreliable, requiring extensive filtering and manual verification.

\textbf{Curated Lists:} Static IoC lists, while easy to use, are rarely sufficient alone. They are typically combined with RegEx or manual validation to boost coverage. Tools such as Twiti \cite{Twiti} and CTI View \cite{CTI_View} use curated lists for labeling or cross-checking, though the lists’ limited update frequency can hinder relevance for emerging threats.

\textbf{Unsupervised Methods:} Techniques like those used in TIMiner \cite{TIMiner} leverage context-based heuristics and keyword co-occurrence to identify IoCs. While scalable, these methods are prone to context-related mislabeling.

We evaluate the most prominent ground truth creation methods and combinations of methods in Section~\ref{subsec:GT_eval}.

\begin{table*}[ht!]
\centering
\renewcommand{\arraystretch}{1.2}
    \resizebox{\linewidth}{!}{%
\begin{tabular}{|c|c|c|c|c|c|c|c|c|c|c|c|c|}
\hline
\multirow{2}{*}{\textbf{Technique}} & \textbf{IoC extraction} & \multicolumn{6}{c|}{\textbf{Ground Truth (GT) Creation Methods}} & \multicolumn{2}{c|}{\textbf{GT Size}} & \multicolumn{2}{c|}{\textbf{Availability}} & \textbf{Code}\\ \cline{3-12}
&  \textbf{Tool} & \textbf{Manual} & \textbf{VT} & \textbf{RegEx} & \textbf{TI Platforms} & \textbf{Lists} & \textbf{Unsupervised} & \textbf{Malicious} & \textbf{Not Malicious} & \textbf{Code} & \textbf{Dataset} & \textbf{Usability}\\
\hline
\multirow{2}{*}{Rule Based} & IoC Searcher \cite{GoodFATR} & \fulldot & & & & & & 29 & 77 & \cmark & \cmark & \fulldot \\
& STIXnet \cite{STIXnet} & \fulldot & & & & & & & & \cmark &   & \fulldot \\
\hline
\multirow{3}{*}{NLP} & ChainSmith \cite{ChainSmith} & \fulldot & & & & & & 6,264 & &   &   &  \\
& TIMiner \cite{TIMiner} & & & \cmark & & & \cmark & & &   &   &  \\
& Xiao \cite{Two-phase_System} & -- & & & & & & 435 & &   &   &  \\
\hline
\multirow{4}{*}{Machine Learning} 
& IoCMiner \cite{IoCMiner} & \emptydot & \cmark & & & & & 45 & 297 & \cmark & \cmark  & \fulldot \\
& Gharibshah \cite{Extracting_actionable_information} & -- & \cmark & & & & & 14,268 & &   &   &  \\
& Twiti \cite{Twiti} & \emptydot & \cmark & & & \cmark & & 50,653 & & \cmark &   &   \\
& IoC Stalker \cite{IoCStalker} & & \cmark & \cmark & & & & 63,903 & 439,586 & \cmark &   & \halfdot \\
\hline
\multirow{7}{*}{\shortstack{Deep Learning \\ \& LLM}} 
& Long et al. \cite{Collecting_Indicators} & -- & & & & & & 69,032 &  &   &   &   \\
& iACE \cite{iACE} & \fulldot & & \cmark & \cmark & & & 1,500 & 3,000   &   &   &   \\
& AITI \cite{AITI} & -- & & & & & & 1,782 & 1,782   &   &   &  \\
& AspIOC \cite{AspIoC} & & \cmark & \cmark & & & & 50,235 & 50,185  & \cmark &   &  \\
& STIOCS \cite{STIOCS} & -- & & \cmark & \cmark & & & 5,922 & & \cmark &   &   \\
& Zhou et al. \cite{Automatic_Identification}& -- & & & & & & 69,032 & &   &   &   \\
& CTI View \cite{CTI_View}& -- & & \cmark & \cmark & \cmark & & 17,364 & &   &   &   \\
\hline
\multirow{2}{*}{Graph-based DL} & HinCTI \cite{HinCTI} & \halfdot & \cmark & & \cmark & & & 11,340 & &   &   &   \\
& HINTI \cite{HINTI} & -- & & & & & & 30,000 & & \cmark &   &   \\
\hline
\end{tabular}
}
\caption{Summary of IoC extraction tools and their ground truth (GT) creation methods, grouped by extraction technique. The \textit{Manual} column denotes the extent of manual labeling, with {\fulldot} indicating extensive and clearly documented manual labeling, {\halfdot} partial manual labeling with documentation, {\emptydot} minimal manual labeling with documentation, and ``{--}'' a claim of manual labeling but without sufficient explanation. Check marks (\cmark) in the \textit{GT Creation Methods} columns indicate that the paper employed the corresponding method. We report malicious and non-malicious samples' GT sizes where available.  {\cmark} in \textit{Availability} indicates publicly available code and/or dataset. A {\fulldot} in the \textit{Usability} column marks that we successfully used the tool, while a {\halfdot} indicates that the tool was usable but out of scope (e.g., works on input types that do not include threat reports).}

\label{tab:ioc_tools}
\end{table*}

\subsection{Investigating Errors of Existing Ground Truth Creation Systems}
\label{subsec:Errors}

To understand the errors that common ground truth creation methods make and their sources, we review two representative systems that are often used to collect IoCs found in unstructured threat reports, IoC Searcher~\cite{GoodFATR} and AlienVault~\cite{iACE,AlienVault}. We evaluate the tools on 10 threat reports published in 2024, sourced from AlienVault’s database, and parsed manually by two analysts, resulting in 517 confirmed IoCs. IoC Searcher is a state-of-the-art rule-based tool with public code that has been compared to other similar tools by previous work \cite{GoodFATR,IoCStalker}. The following evaluation provides a robust baseline for assessing both tools.

IoC Searcher produced 126 false positives. The most common error was ``Not Malicious'' (52 instances), where benign entities like domains (e.g., \texttt{msn.com}, \texttt{github.com}), IPs, (e.g., \texttt{1.1.1.1}, \texttt{4.2.2.4}), URLs, and hashes were misclassified as IoCs solely based on their format, without considering the surrounding context indicating them as benign. Next came "File Path," a semantic misclassification where benign paths were flagged as domains or URLs due to shared string patterns (e.g., \texttt{do.zip}, \texttt{asp.net}). Other common errors included \texttt{TLD} misclassification as IoC domains, indicators pointing to the ``Reporting Company'' extracted as IoCs, partial substrings extraction of the full IoC, and extracting indicators with placeholders or redacted labels(e.g., \texttt{www.[redacted].com}, microsoft-update-com.github.io/XXX/update.html?id=[GUID ). These results reflect the limits of these tools in contextual understanding and formatting variability.
Despite these issues, IoC Searcher showed high recall, missing only nine indicators. After manually investigating the misclassified IoCs, we found that most errors stemmed from newlines splitting indicators, incorrect type detection, or partial matches (``Found Longer Variant'').

Most of the 64 false positives in AlienVault’s dataset were indicators that did not appear in the reports, often due to inferred relationships like alternate file hashes (e.g., SHA1$\rightarrow$SHA256). Even though these indicators might be useful, they do not appear in the report and are difficult to validate.
In terms of false negatives, the dataset missed 67 IoCs. Among these, 60 were entirely missing, six were split across lines, and one was only partially matched. These results highlight that even widely used platforms like AlienVault suffer from coverage gaps and labeling inconsistencies.

Overall, this analysis illustrates the key limitations of automated IoC extraction tools: poor context handling, formatting brittleness, and inconsistent labeling. These challenges motivate our hybrid human-in-the-loop (HITL) design, which combines automated context-aware suggestions with validation from analysts, thus avoiding these pitfalls and producing a more precise and high-quality dataset.

\section{IoC Labeling Methodology}
\label{sec:Methodology}

In this section, we present our methodology for labeling IoCs extracted from unstructured threat reports. We begin by formally defining the problem, the goals of our methodology, and the constraints that guide our human-in-the-loop design. We then describe LANCE, our LLM-based system for generating explainable IoC labels, detailing its architecture. Next, we introduce the custom user interface developed to support analyst interaction, enabling efficient validation and correction of LLM-generated labels. Finally, we explain the full dataset creation process.

{We limited our scope to the four most common and reliably verifiable indicator types, IP addresses, domain names, URLs, and file hashes, which are well-supported by VirusTotal and commonly shared in structured platforms such as AlienVault. While other indicators (e.g., Bitcoin addresses, email addresses, file paths, registry keys) are indeed relevant, they are harder to verify at scale using current automated methods, and less consistently present across reports. We leave the inclusion of such indicators to future work.}

\subsection{Problem Definition, Goals, and Constraints}
\label{subsec:GoalsAndConstraints}

In Section~\ref{subsec:GTgenerationMethods}, we examined the ground truth creation methodology of 18 prominent IoC extraction papers, analyzing the transparency of their methods, as well as the availability, and usability of their code and datasets, as summarized in Table~\ref{tab:ioc_tools}. We observed significant variance in how these works created ground truth. Manual labeling was often poorly documented, making it unclear what role human input played in the dataset creation process. Furthermore, only one paper made both its code and dataset publicly available, with one more sharing them upon request. These limitations have been echoed in recent studies~\cite{Reproducability, GoodFATR, STIOCS, ChainSmith, Survey1}.

The lack of transparency in ground truth construction, combined with the unavailability of the underlying datasets, severely impedes reproducibility, fair tool evaluation, and comparison. This not only slows progress in the field but also diminishes trust in automated IoC extraction systems due to unclear validation paths and opaque data pipelines.

To overcome these limitations, we introduce a novel IoC extraction and labeling methodology based on a human-in-the-loop (HITL) framework. We then use to construct the first openly available and thoroughly documented benchmark dataset for IoC extraction from threat reports, called \textbf{PRISM}. Our HITL pipeline integrates large language model (LLM)–generated labels with analyst oversight, enabling both high-precision labeling and significant reductions in human effort. Rather than treating human labeling and automation as mutually exclusive approaches, we combine them into a cohesive workflow in which explainable LLM predictions guide human decisions, and human validation resolves ambiguity.

Our methodology aims to achieve these main objectives:

\begin{itemize}
    \item \textbf{High Accuracy:} The dataset must minimize both false positives and false negatives to ensure fidelity.

    \item \textbf{Manual Work Factor Effectiveness:} The labeling workflow should reduce total analyst hours without compromising label quality.

\end{itemize}

{Throughout this work, we distinguish between \textit{labeling}, which refers to assigning IoC/nonIoC status to individual indicators, and \textit{ground truth generation}, which refers to the full process of creating high-quality labels through labeling, human validation, and dispute resolution.}

\subsection{LANCE}
\label{subsec:AuyomatedIoCextractionMethodology}

\begin{figure}
    \centering
    \includegraphics[width=\linewidth]{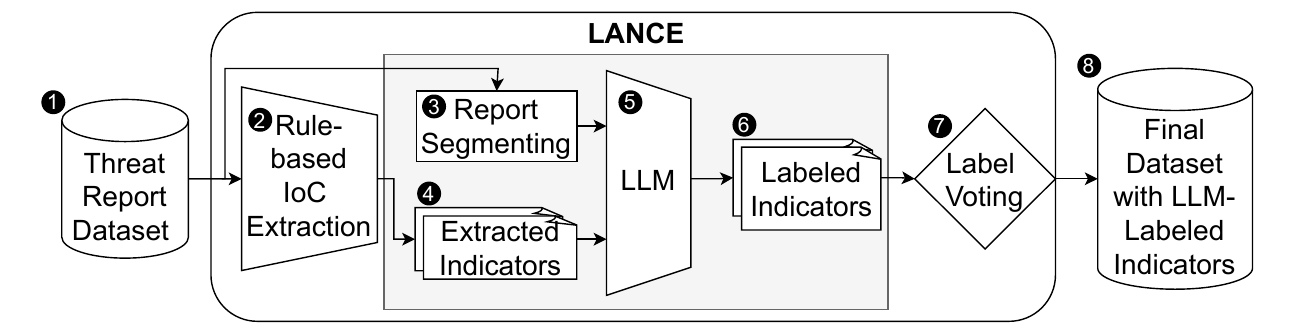}
    \caption{Overview of the LANCE pipeline. Indicators are extracted using regular expressions, labeled by an LLM using contextual report segments, and finalized through a voting mechanism to resolve overlapping predictions.}

    \label{fig:LLM_pipeline}
\end{figure}

To streamline the IoC extraction and labeling process, so we can minimize the time the analysts need to spend on the task, we developed LANCE: LLM-Assisted Notation and Classification Engine, a tool that automatically extracts and labels Indicators of Compromise (IoCs) from unstructured threat reports. An overview of LANCE can be seen in Figure \ref{fig:LLM_pipeline}. The extraction phase utilizes a rule-based approach using regular expressions, while a Large Language Model (LLM) performs the labeling. Explainability is central to our HITL design. By requiring the LLM to justify each classification decision, we shift the downstream analyst’s role from being a manual extractor to that of an informed validator.

\subsubsection{\textbf{Rule-based Extraction}}
Regular expression-based tools are highly effective in achieving high recall for IoC extraction, but struggle with precision. As discussed in Section~\ref{subsec:Errors}, they fail to differentiate between malicious and non-malicious indicators. Given the above, we employed regular expressions for the extraction step to ensure broad coverage of potential IoCs.

While RegEx-based tools like IoC Searcher maximize recall, they lack contextual understanding. In our pipeline, this extraction phase is intentionally broad: it ensures full coverage so that no potential IoC is missed, with human and LLM refinement steps later correcting for over-inclusion. This approach supports a recall-first strategy where precision is restored through layered validation.

For this purpose, we adopted IoC Searcher, a regular expression-based tool proposed by Caballero et al.\cite{GoodFATR}, and integrated it into our pipeline (Figure \ref{fig:LLM_pipeline} part 2). As discussed in Section~\ref{subsec:Errors}, IoC Searcher demonstrates high recall across various IoC types, making it well-suited for the initial extraction phase \cite{GoodFATR, IoCStalker}. We used the tool without modifications, as it has been evaluated in prior research and outperformed other regular expression-based IoC extraction tools \cite{GoodFATR}.

\subsubsection{\textbf{LLM-based Indicator Labeling}}
\label{subsec:LLMmethodology}

For the labeling phase, we are employing a conversational LLM with zero-shot learning. Each report is processed iteratively for each IoC type, with the LLM classifying extracted indicators as either \textit{IoC} or \textit{nonIoC} based on their surrounding context.

To handle long reports, we implemented a rolling window approach with a segment size of approximately 8000 characters (Figure \ref{fig:LLM_pipeline} part 3), ensuring splits occurred at the nearest whitespace to preserve readability. This approach has been shown to improve the performance of LLM \cite{DivideNConquer,LLMxMapReduce}, {allowing LANCE to handle reports of arbitrary length}. 
To mitigate content loss at segment boundaries, we introduced a 50\% overlap between consecutive segments. This implementation ensured that no critical context was lost in segmentation \cite{LLMOverlap}.

For each segment, we provide the LLM with the segment text along with a list of all extracted indicators (of the target type) from that segment (Figure \ref{fig:LLM_pipeline} parts 3 and 4). We then prompt it to classify each indicator as IoC or nonIoC and provide a justification for its label (Figure \ref{fig:LLM_pipeline} part 5). Including justifications enhances the labeling process's interpretability, offering insights into why a particular indicator was classified as malicious or benign. This level of explainability is crucial for understanding the decision-making process and enables better auditing of the output.

Since some indicators appeared in multiple segments, we implemented a voting mechanism to consolidate classifications across the entire report (Figure \ref{fig:LLM_pipeline} part 7). The voting thresholds and the ratio of \textit{IoC} to \textit{nonIoC} labels required for a final classification were treated as hyperparameters. We empirically fine-tune the voting threshold for each type (see Appendix~\ref{app:LANCEPrompts}).

Regarding the prompts given to the LLM, after extensive iteration and error analysis,
we finalized a robust structure that improves both precision and model reliability across IoC types. The final prompts consist of the following components:
\begin{itemize}
    \item Specify the role of the LLM.
    \item Clear definition of what the input is.
    \item Clear definition of the classification task.
    \item Explicit definitions for ``IoC'' vs ``nonIoC''.
    \item Descriptions of common false positives/negatives.
    \item Clear definition of the justification task.
    \item Explicit instructions to be careful and thorough.
    \item Specification for the expected output format.

\end{itemize}

These design elements significantly reduced error rates, particularly in complex cases, including domains and URLs, where indicators often resemble benign artifacts or contain embedded nonIoC substrings. We provide a detailed analysis of the prompt engineering phase and an evaluation of the prompts in Appendix~\ref{app:LANCEPrompts}.

\subsection{Custom User Interface}
\label{subsec:UserInterface}

Once LANCE-labeled IoCs are extracted, they are shown to the analyst in context within the related threat report document, alongside LLM-generated label explanations. This structure reduces human effort
and allows experts to focus only on the most ambiguous cases with full visibility into the model’s reasoning.

To this end, we developed a custom user interface (UI) to assist analysts in the label verification/correction task. In the UI, the analysts are able to load a threat report for processing. All indicators (IPs, URLs, domains, and hashes) extracted by LANCE  are highlighted with a color-coded scheme, making them easily visible to the analysts~\cite{UI1,UI2,UI3}. The colors we use are either green, for non-IoC indicators, or red for indicators labeled as IoC. A view of the User Interface can be seen in Appendix~\ref{app:UI}, Figure~\ref{fig:UI2}.

The user interface provides the analyst with the LANCE-generated label and justification. The analyst has the option to either accept or override the generated label. The interface also enables analysts to view justifications by hovering over indicators as well as adding additional comments for future reference. After the analyst finalizes all labels, they can export them in a structured format along with all justifications and additional comments optionally provided by the analyst.

\begin{figure*}
    \centering
    \includegraphics[width=\textwidth]{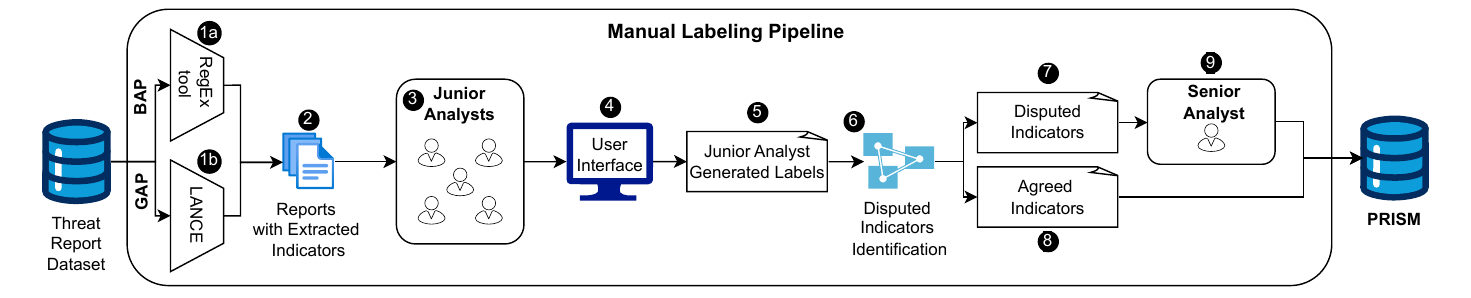}
    \caption{Structure of the manual annotation process used to create the PRISM dataset. The process includes two phases: the Baseline Annotation Pass (BAP), where analysts label indicators without assistance, and the Guided Annotation Pass (GAP), where analysts review indicators pre-labeled by LANCE along with justifications. In both phases, labels from junior analysts are compared, and any disagreements are resolved by a senior analyst. The final consensus labels are incorporated into the PRISM dataset.}
    \label{fig:Experiment}
\end{figure*}

The user interface is not merely a visualization tool. It operationalizes the human-in-the-loop paradigm by making model predictions transparent, actionable, and easy to confirm or override. This design reduces annotation effort and supports our goal of a scalable expert-guided validation.

\section{PRISM Dataset Creation}
\label{subsec:DatasetMethodology}


We now describe how we use the HITL system described earlier (Figure~\ref{fig:TotalPipeline}) to create our PRISM dataset. The labeling pipeline we use is depicted in  Figure~\ref{fig:Experiment}. 
We start by describing our ground truth generation methodology with human vetting, and then discuss our concrete application of this methodology to the PRISM dataset.
We describe PRISM first here, and later (in Section~\ref{sec:Evaluation}), use the manually verified IoC labels obtained during this dataset creation process to evaluate LANCE's accuracy by measuring its level of agreement with expert threat analysts.

\subsection{Methodology Overview}
\label{subsec:DatasetMethodologyOverview}

For the dataset creation, given a set of threat reports, we first divide them into two equal subsets, ensuring a balance in average report length, indicator density, and distribution of indicator types. The two sets are used in two separate manual labeling pipelines: the Baseline Annotation Pass (BAP) and the Guided Annotation Pass (GAP). A detailed analysis of each is given in this section. 

All reports are processed through LANCE to extract and label indicators. For the creation of PRISM, we assigned each report to junior analysts who parsed the report independently to facilitate comparison and identify discrepancies. 

To further increase the quality of the PRISM dataset, as well as to evaluate the effects LANCE had on the junior analysts, we also employed a senior analyst. The senior analyst annotated only the reports that had indicators in which the junior analysts disagreed. This way, we minimized the time the senior analyst needed to spend on labeling indicators and used their expertise only on the difficult-to-label indicators.

\subsubsection{\textbf{Baseline Annotation Pass (BAP)}}

For this part, we gave the first set of reports to the junior analysts, with the goal of labeling the indicators in the reports as ``IoC'' or ``nonIoC''. The reports were given to the analysts through the user interface described in Section~\ref{subsec:UserInterface}, with the crucial difference that the analysts could not see the LANCE-generated labels and justifications. In each report, they were able to see all the indicators highlighted in the report with a neutral color (blue). A view of the User Interface used for BAP can be seen in Appendix~\ref{app:UI} in Figure~\ref{fig:UI1}. The analysts were able to assign and change the label of each indicator in an easy and quick way. 

The junior analysts were instructed not to use any external sources that would help them choose the label of an indicator, and to rely on the context in which the indicator appears in the report. The types of indicators that were targeted were explained to the analysts so they could avoid confusion with indicator-like strings that the regular expression part of LANCE might have extracted. In the cases of an indicator appearing more than once in a report, the label assigned by the analyst to that indicator appeared in all instances of the indicator. 

A short video tutorial of the User Interface was provided, as well as written user instructions and task details. 
Finally, after labeling all the indicators of a report, the labels were stored in a JSON file.

The junior analyst received the following instructions:
\begin{itemize}
    \item Label all extracted indicators (domain, IP, URL, HASH) in each of the assigned reports as \textbf{IoC} or \textbf{nonIoC} using the web interface.
    \item An \textbf{IoC} is an indicator that suggests compromise or malicious activity based on the context provided in the report.
    \item Complete labeling for each report in \textbf{one sitting}, but you may take breaks between reports.
    \item \textbf{Do not} use the internet or external resources. Instead, base your decisions only on the report content.
\end{itemize}

After the end of the junior analysts' task for BAP, we used the labels from the junior analysts to find \textit{disputed} indicators. We define a \textit{disputed} indicator as an indicator for which there was no total agreement between the labels of the analysts and the LANCE-generated label. For example, if LANCE and a junior analyst agree that an indicator should be labeled as an ``IoC'' but another junior analyst labeled that indicator as ``nonIoC'', then this is a \textit{disputed} indicator.

\begin{figure*}
    \resizebox{\textwidth}{!}{%
    \centering
    \includegraphics[width=\textwidth]{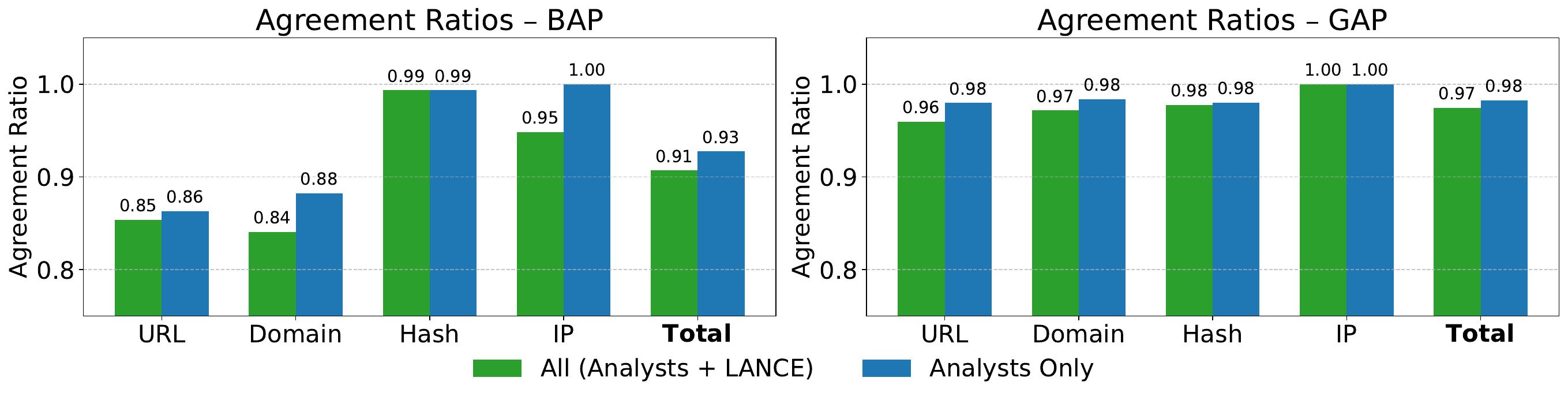}
    }
    \caption{Agreement ratios across IoC types during the manual annotation process. The left plot shows results from BAP (Baseline
    Annotation Pass) and the right plot from GAP (Guided Annotation Pass). Each bar pair reflects agreement between the junior
    analysts (Analysts Only) and including LANCE (All).}
    \label{fig:Agreement}
\end{figure*}

The reports that included at least one disputed indicator were tagged as disputed reports and were given to a senior analyst to determine the correct label of the indicators. These reports were given to the senior analyst through the same UI and with the same instructions given to the junior analysts.

The senior analyst was tasked to label all indicators in the disputed reports, not only the disputed indicators. This way, we could see if other indicators that were labeled falsely by both the junior analysts and the LLM.

\subsubsection{\textbf{Guided Annotation Pass (GAP)}}

For the second part of the ground truth creation setup, we used the second set of reports. The setup for GAP was similar to that of BAP, with the exception that the analysts did have access to the labels that were assigned to the indicators by LANCE, as well as the justifications that LANCE provided for this choice. This allowed the junior analysts to be ``advised'' by LANCE while labeling each indicator. 

The junior analysts did not have access to any source of information other than the report and LANCE-generated labels and justifications. 

The junior analyst received the following instructions in this part:
\begin{itemize}
    \item Review the assigned reports where indicators (domain, IP, URL, HASH) have already been labeled by the system, along with justifications.
    \item An \textbf{IoC} is an indicator that suggests compromise or malicious activity based on the context provided in the report.
    \item Identify and correct any incorrect labels, providing a short comment for corrections where necessary.
    \item Complete labeling for each report in \textbf{one sitting}, but you may take breaks between reports.
\end{itemize}

The main motivation for this part of the setup was to evaluate the influence the LANCE labels and justifications had on the analysts.  Given that the LLM might have more information about certain indicators, we expected it to be able to ``persuade'' the analysts that its label was correct without changing their minds when its approach is incorrect.  

After annotation, disputed indicators were identified, and the reports that contained at least one disputed indicator were assigned to the senior analyst for final validation.

\subsection{PRISM Dataset Details}
\label{subsec:DatasetCreationMethodology}

To construct \textbf{PRISM}, our benchmark dataset of labeled IoCs, we first curated a diverse set of 500 threat reports from ORKL~\cite{ORKL}, an online repository containing over 13,000 reports from 11 reputable sources, including MITRE, APTnotes, CyberMonitor, and Malpedia~\cite{mitre, APTnotes, CyberMonitor, malpedia}. These reports span a 20-month period from April 2023 to November 2024 and vary in length, origin, and IoC density.

To ensure relevance and quality, we filtered for reports that also appeared in AlienVault~\cite{AlienVault, iACE}, a widely used threat intelligence platform. Specifically, we selected reports published by AlienVault's official user account, which aggregates high-quality reports from vendors such as Kaspersky, Palo Alto Networks, and Microsoft. From this filtered corpus, we identified 50 reports that (i) appeared on both ORKL and AlienVault, (ii) were fewer than 30 pages in length, and (iii) contained more than five unique indicators. Although this threshold provides only an upper bound on the number of true IoCs, it serves as an effective proxy for report richness. {The 30-page limit was a practical choice to reduce the workload of human annotation during dataset creation.}

These 50 reports were used to construct the ground truth dataset following the annotation methodology described in Section~\ref{subsec:DatasetMethodologyOverview}. To annotate the reports, we recruited five junior analysts-PhD students specializing in cybersecurity, and one senior analyst with seven years of experience in threat hunting.

The 50 reports were divided into two equal subsets of 25 for the two annotation phases: the \textit{Baseline Annotation Pass (BAP)} and the \textit{Guided Annotation Pass (GAP)}. The sets were balanced in terms of average report length, indicator density, and distribution of IoC types. Table~\ref{tab:total_indicators} shows the total number of non-unique indicator instances per type for each set. 

For the LANCE implementation, we used ChatGPT 4o as the underlying LLM \cite{GPT4o}. Each report was independently labeled by two junior analysts. We evenly distributed the workloads in terms of reports, indicator counts, and inter-annotator overlap across the analyst pairs.

\begin{table}[h!]
\centering
    \resizebox{0.7\linewidth}{!}{%
\begin{tabular}{|l|c|c|c|}
\hline
\textbf{Indicator Type} & \textbf{BAP} & \textbf{GAP}& \textbf{Total} \\
\hline
IP      & 177 & 112  & 289  \\
Domain  & 694 & 729  & 1423 \\
URL     & 426 & 445  & 871  \\
HASH    & 962 & 758  & 1720 \\
\hline
\textbf{Total} & 2259 & 2044 & 4303 \\
\hline
\end{tabular}
}
\caption{Analysis of non-unique indicator instances in reports for BAP and GAP, and in total}
\label{tab:total_indicators}
\end{table}

Disagreements between annotators, or between annotators and the LANCE-generated labels, were flagged as \textit{disputed}. We identified 15  reports from BAP that had at least one \textit{disputed} indicator and nine from GAP. These reports were escalated to the senior analyst, who re-labeled all indicators in those reports to ensure consistency and completeness. Figure~\ref{fig:Agreement} shows the agreement ratios between annotators and with LANCE.



\begin{table}[h!]
\centering
    \resizebox{\linewidth}{!}{%
\begin{tabular}{|l|cc|cc|c|}
\hline
\textbf{Indicator} & \multicolumn{2}{c|}{\textbf{BAP}} & \multicolumn{2}{c|}{\textbf{GAP}}& \textbf{Total} \\
\cline{2-5}
\textbf{Type} & \textbf{IoC} & \textbf{nonIoC}  & \textbf{IoC} & \textbf{nonIoC} & \textbf{(IoC + nonIoC)} \\
\hline
IP      & 97  & 0   & 58  & 3   & 158  \\
Domain  & 176 & 113 & 105 & 143 & 537  \\
URL     & 156 & 63  & 100 & 47  & 366  \\
HASH    & 309 & 1   & 400 & 3   & 713  \\
\hline
\textbf{Total} & 738 & 177 & 663 & 196 & 1774 \\
\hline
\end{tabular}
}
\caption{Analysis of unique IoC and nonIoC indicators for BAP and GAP, including total sums and per IoC type.}
\label{tab:total_unique_in_GT}
\end{table}

This methodology resulted in 1,774 high-confidence, expert-reviewed labels. The combination of unaided annotation (BAP) and LANCE-assisted annotation (GAP), along with senior analyst arbitration, ensures PRISM is both rigorous and reproducible, offering a high-quality ground truth for fair evaluation and training of IoC extraction systems. Table~\ref{tab:total_unique_in_GT} presents the number of unique labeled indicators, both IoC and nonIoC, for each annotation phase and indicator type.

\begin{figure*}
    \centering
    \includegraphics[width=\textwidth]{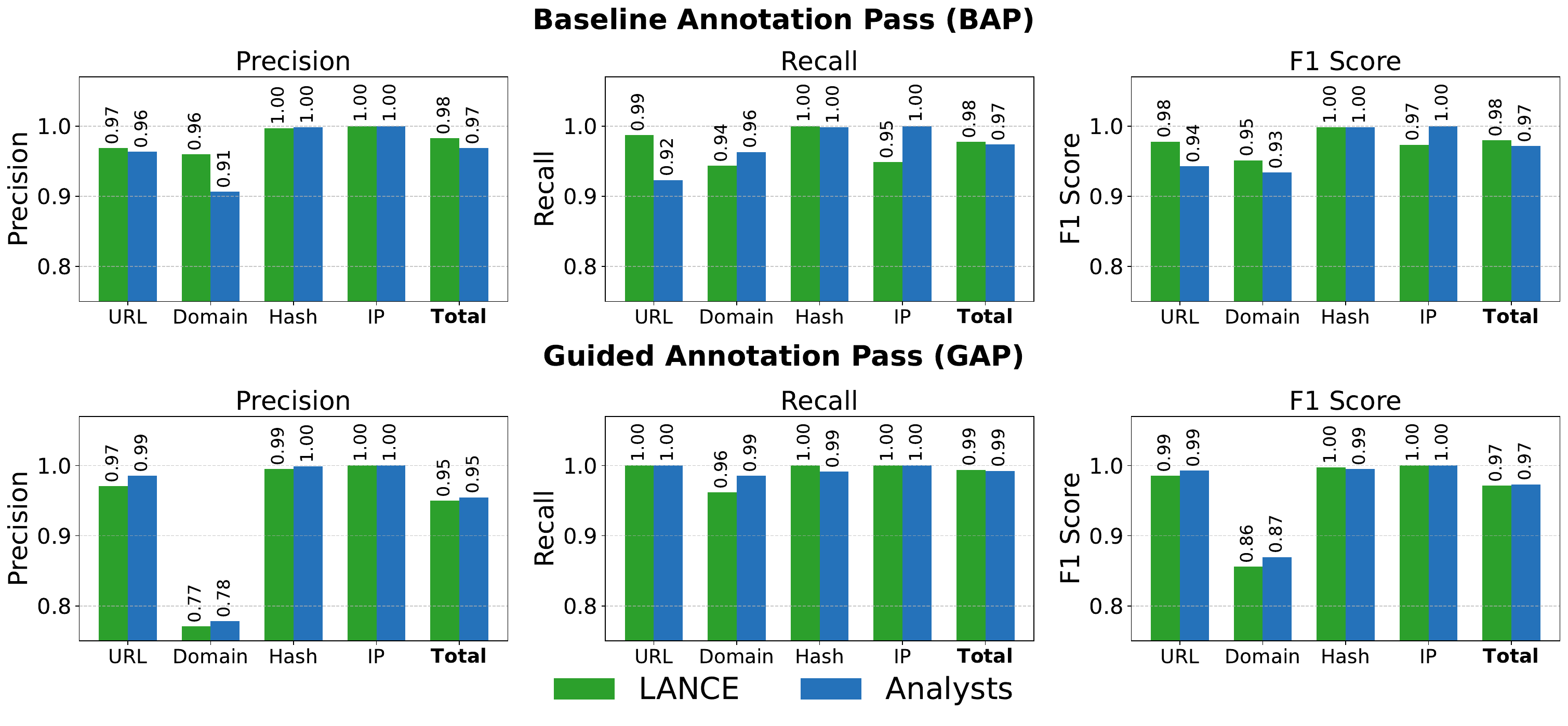}

    \caption{Comparison of LANCE and analyst performance across precision, recall, and F1 score during both annotation phases. The top row shows results from BAP, while the bottom row shows GAP. }

    \label{fig:gpt_vs_analyst_perf}
\end{figure*}

The Precision, Recall, and F1 scores that the junior analysts achieved compared to the final ground truth can be seen in Figure~\ref{fig:gpt_vs_analyst_perf}.

\section{IoC Labeling Evaluation}
\label{sec:Evaluation}
In this section, we evaluate LANCE by comparing it with other prominent automated ground truth creation methods. We also evaluate its downstream utility and its ability to generalize across different state-of-the-art LLMs. Additionally, we evaluate the proposed HITL pipeline by comparing its impact on junior analysts, specifically in terms of how it influenced their precision, recall, and labeling speed. Finally, we discuss some common cases of indicators that were misclassified by LANCE.

\subsection{LANCE Evaluation}

\subsubsection{\textbf{Comparison with Existing Automated IoC Labeling Systems}}
\label{subsec:GT_eval}

We use PRISM, our dataset of IoCs that are manually labeled by multiple threat analysts (see Section~\ref{subsec:DatasetMethodology}), to evaluate the most prominent existing automated ground truth (GT) creation methods discussed in Section~\ref{subsec:GTgenerationMethods}, and compare them with LANCE. Specifically, we assess four strategies based on their extraction and labeling performance: (1) IoC Searcher with whitelist-based filtering~\cite{GoodFATR}, (2) AlienVault~\cite{iACE}, and (3,4) VirusTotal-based labeling with two thresholds~\cite{AspIoC, HinCTI, Extracting_actionable_information, IoCStalker, Twiti}. 
None of the mentioned GT creation methods require training.

\textbf{Evaluation Setup:}
For the RegEx + Whitelist method, we implemented the GT creation pipeline from the GoodFATR paper~\cite{GoodFATR}, which supplements rule-based extraction by IoC Searcher with whitelist and frequency-based filtering to reduce false positives. The AlienVault method involved retrieving all IoCs (including inactive ones) linked to each report from the AlienVault platform \cite{iACE}. All indicators not in the platform were considered unlabeled and treated like nonIoCs. For the VirusTotal-based methods, we followed prior work~\cite{IoCStalker,Twiti,AspIoC,Extracting_actionable_information} by extracting indicators with IoC Searcher and labeling them based on VirusTotal lookups. We used two thresholds: 
\textit{Threshold 1} labels an indicator as IoC if at least one vendor flags it as malicious, and as nonIoC if all vendors deem it benign.
\textit{Threshold 5} raises the bar, requiring five or more malicious detections to label an indicator as IoC, while indicators with fewer than five malicious detections are labeled nonIoC. 
Indicators not included in the database are left unlabeled and are considered as nonIoC.

\begin{figure}
    \centering
    \includegraphics[width=\linewidth]{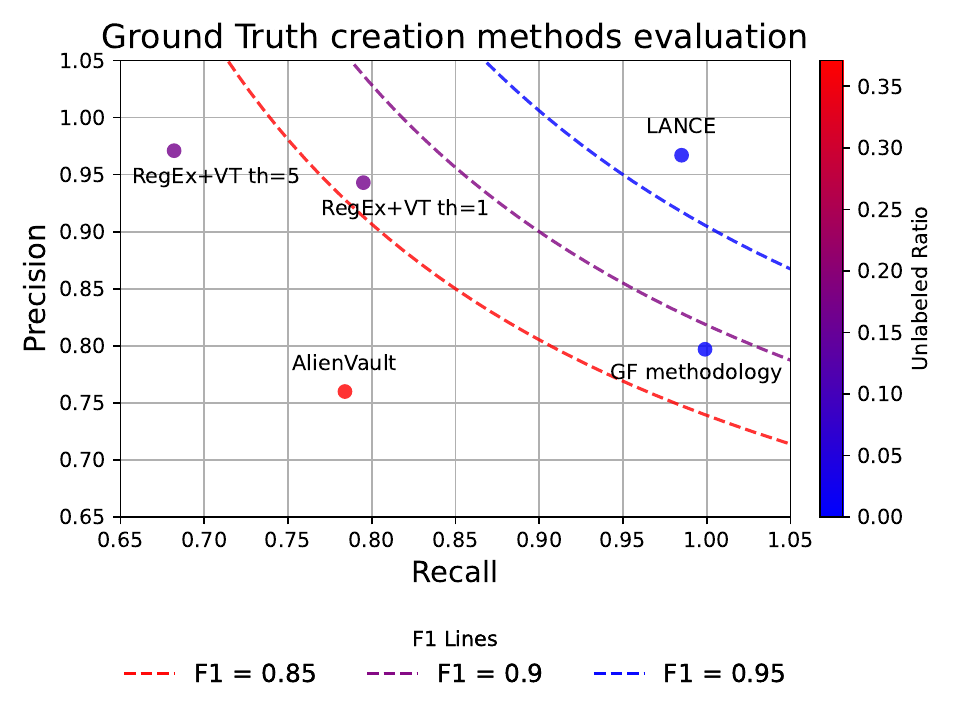}

    \caption{Performance evaluation of prominent Automated ground truth creation methods}

    \label{fig:GT_method_eval_total}
\end{figure}

\textbf{Coverage Analysis:}
We assume a total of 1,789 candidate indicators, extracted using IoC Searcher, a state-of-the-art rule-based tool~\cite{GoodFATR, IoCStalker, AspIoC, Extracting_actionable_information}. 
Figure~\ref{fig:GT_method_eval_total} shows the ratio of unlabeled indicators across methods. The RegEx + Whitelist method labels all extracted indicators but still suffers from poor precision due to insufficient context awareness. The AlienVault method yields the most unlabeled indicators ($37\%$). When evaluating the ground truth provided by AlienVault, we also observe that the sum of Labeled and Unlabeled indicators exceeds the total of 1,789. This can be explained by the analysis in Section~\ref{subsec:Errors}. The VirusTotal-based methods leave over 280 indicators unlabeled, suggesting that $16\%$ of the indicators are absent from the VirusTotal database. In contrast, LANCE labeled over $99\%$ of all extracted indicators, with the few unlabeled cases attributed to malformed LLM outputs.

\begin{tcolorbox}[boxsep=-0.5mm, boxrule=0.4mm]
\textbf{Takeaway:} LANCE provides more comprehensive extraction and labeling coverage than other prominent automated ground truth creation methods.
\end{tcolorbox}

\textbf{Performance Analysis:}
Figure~\ref{fig:GT_method_eval_total} summarizes overall precision, recall, and F1 score for each method. 
We further break down performance across the four IoC types (IP, hash, URL, domain) in Appendix~\ref{app:GTevalDetails} (Table~\ref{tab:GT_creation_method_eval_on_GT}). 

The RegEx + Whitelist method achieves almost perfect recall but low precision, particularly for domains and URLs, making it unsuitable for high-fidelity ground truth creation due to excessive false positives. AlienVault shows the lowest recall for URLs ($25\%$) and significant gaps in domain coverage ($73\%$ recall), likely due to incomplete extraction. Its precision on hashes is also low ($69\%$) because many labeled hashes do not appear in the report text.
VirusTotal with threshold 1 performs reasonably well, achieving an average F1 score of $86\%$, with consistent scores across indicator types. Threshold 5 has increased precision; however, it suffers from low recall due to strict filtering criteria, reducing its reliability for comprehensive dataset creation.

LANCE outperforms all other methods, consistently achieving over $90\%$ F1 score across all types and $97.6\%$ overall. Its lowest score, $86.8\%$ recall on domains, is due to intentional disagreements between LANCE's labels and the senior analyst in cases involving compromised legitimate websites. Notably, LANCE’s justifications showed awareness of this nuance, reinforcing the importance of contextual labeling aligned with downstream use cases. {We further analyze mislabeled examples in Section~\ref{misclassifications}.}
These discrepancies between LANCE and the senior analyst's labels highlight an important challenge: the definition of what constitutes an IoC can, in some cases, be ambiguous as it may depend on factors such as the time of observation, the current threat landscape, and the specific downstream task where the IoC labels will be used (e.g., detection, attribution, or sharing)\cite{IoCStalker,IoCdef1,IoCdef2}. This further justifies the need for manual verification with a human-in-the-loop approach.

\begin{tcolorbox}[boxsep=-0.5mm, boxrule=0.4mm]
\textbf{Takeaway:} LANCE outperforms all other 
prominent automated ground truth creation methods in total 
F1 score, maintaining consistently high performance 
across all IoC types. This makes it the most suitable 
approach for creating ground truth in high-stakes 
security contexts.
\end{tcolorbox}

\subsubsection{\textbf{LANCE Comparison With Naive LLM Prompts}}
\label{subsec:NaiveGPT}

We evaluate how the LANCE pipeline impacts labeling performance by comparing ChatGPT-4o \cite{GPT4o} with and without it.
For this experiment, we input to ChatGPT the 50 reports included in PRISM and naively prompt it to extract all IoCs from the text. We run four such experiments, one for each type of indicator. Finally, we compare the extraction and labeling results to LANCE. 
As mentioned in Section~\ref{sec:Methodology}, we implement zero-shot learning, so there is no training needed.

\begin{figure}
    \centering
    \includegraphics[width=\linewidth]{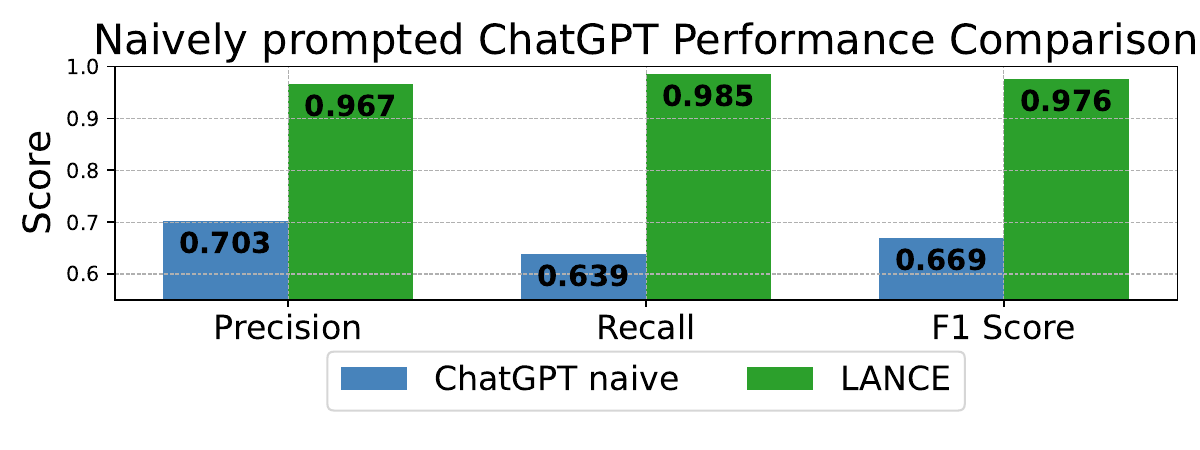}
    \caption{Performance comparison of LANCE and baseline, naively prompted ChatGPT on PRISM.}
    \label{fig:GPT_naive}
\end{figure}

As seen in Figure~\ref{fig:GPT_naive}, LANCE performs significantly better than the baseline naive ChatGPT in all metrics. Specifically, the baseline approach achieves a $0.669$ F1 score in total, with its F1 scores in URLs and Domains being $0.343$ and $0.517$, respectively. It is evident that the naive approach suffers both in extraction and labeling capabilities, which has also been shown in previous work \cite{naiveLLM}. This demonstrates the necessity of the LANCE approach.

These results show that LANCE is critical in the pipeline. 
Rather than relying on naive prompting, LANCE strategically orchestrates LLM capabilities via prompt design and context-aware segmentation and achieves high precision on complex threat data. Without this pipeline, the LLM underperforms by $30\%$ in F1 score (as shown in Figure~\ref{fig:GPT_naive}).

\begin{tcolorbox}[boxsep=-0.5mm, boxrule=0.4mm]
\textbf{Takeaway:} Without structured guidance, LLMs like ChatGPT-4o underperform in IoC extraction and labeling. LANCE significantly boosts both precision and recall, underscoring the importance of a tailored pipeline.
\end{tcolorbox}

\subsubsection{\textbf{Downstream Utility Evaluation}}
To evaluate the downstream utility of a LANCE-generated dataset, we use LANCE-generated labels to train a downstream IoC-extraction tool. For this evaluation, we chose IoCMiner~\cite{IoCMiner} as the downstream IoC-extraction tool, as it was the only tool in Table~\ref{tab:ioc_tools} that requires training and had usable code. IoCMine~\cite{IoCMiner} is a Machine Learning-based IoC classification tool developed for labeling IPs, URLs, and hashes from tweets. 
{The format of the training dataset provided by the authors of IoCMiner is based on one-hot encoded bag-of-words representations, thus making the reconstruction of the original sentences infeasible.
For this reason, we generate our datasets for this evaluation, using 450 reports that are separate and do not overlap with the 50 reports in PRISM. We use these reports to generate three different datasets, one with LANCE-generated labels and two generated by VirusTotal-based labels (with thresholds 1 and 5). We then use the three different datasets to train three different instances of the IoCMiner model and evaluate each of them on PRISM.}

\begin{figure}
    \centering
    \includegraphics[width=0.8\linewidth]{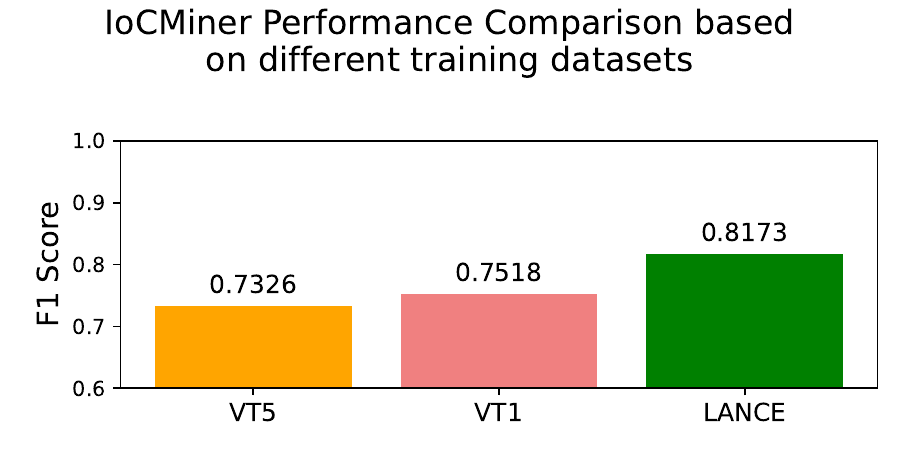}
    \caption{Comparison of IoCMiner's~\cite{IoCMiner} accuracy on PRISM, based on differently created ground truths. The ground truths for this experiment were labeled by VirusTotal with threshold=5 (VT5), VirusTotal with threshold=1 (VT1), and LANCE. We see that the model trained with the LANCE-generated ground truth outperforms the other two.}
    \label{fig:IoCMiner}
\end{figure}

Specifically, to extract English sentences from the reports, we first split the text of the report on periods and then use IoC Searcher \cite{GoodFATR} to see if an indicator of the targeted type is included in each sentence. For the 450 reports of the training set, we generate labels for the indicator in each sentence using the two highest performing ground truth generation methods (VirusTotal with threshold=1 and threshold=5) as indicated in Figure~\ref{fig:GT_method_eval_total}, and LANCE, automatically generating three different training sets (VT1, VT5, and LANCE). We train IoCMiner \cite{IoCMiner} once for each different training set using the available code and without altering the hyperparameters, and evaluate the three separately trained instances of the model on PRISM. The result of the evaluation can be seen in Figure~\ref{fig:IoCMiner}. 

We observe that the model trained on the dataset labeled by LANCE achieves a superior F1 score in PRISM than the models trained on datasets labeled by the other two methods. This is consistent with previous evaluations of LANCE, as in Section~\ref{subsec:GT_eval} we see that LANCE has superior labeling capabilities than Virsus Total with thresholds 1 and 5 and better ground truth enables better-performing models \cite{betterGT1,betterGT2}. More specifically, we see that the LANCE-trained model outperforms the VT1-trained model by over $6\%$ and the VT5-trained model by over $8\%$ in terms of F1 score.

\begin{tcolorbox}[boxsep=-0.5mm, boxrule=0.4mm]
\textbf{Takeaway:} Datasets created by the LLM-based pipeline (LANCE) promote better and more robust training than other automated ground truth creation methods.
\end{tcolorbox}

\subsubsection{\textbf{Generalization Across LLMs}}
\label{PipelineGeneralization}

To test the generalizability of our pipeline, we evaluate it across several state-of-the-art LLMs. This not only tests the adaptability of LANCE but also our broader hypothesis: that well-engineered systems can remain effective across model families by abstracting the labeling logic into prompts and input structure.

We tested LANCE on four additional prominent LLMs: LLama 3.3 70b \cite{LLama3.3}, Gemma 3 27b\cite{gemma3}, Gemini 2.0 Flash \cite{gemini2flash}, and Nvidia Llama 3.1 Nemotron 70b \cite{nvidiallama}. We test those LLMs on PRISM ground truth with the same prompts that we developed in Section~\ref{subsec:AuyomatedIoCextractionMethodology}. For our experiments on GPT and Gemini, we used the provided APIs, while we used the open-source pre-trained models LLama, Nemotron, and Gemma that are available, with quantization, on 2 A40 GPUs with 46 GB of memory each. 
To ensure a fair evaluation, we only tested the LLMs on the 25 reports of BAP (where the GPT-based LANCE labels were not available to analysts during ground truth creation), eliminating potential bias.

\begin{figure}
    \centering
    \includegraphics[width=\linewidth]{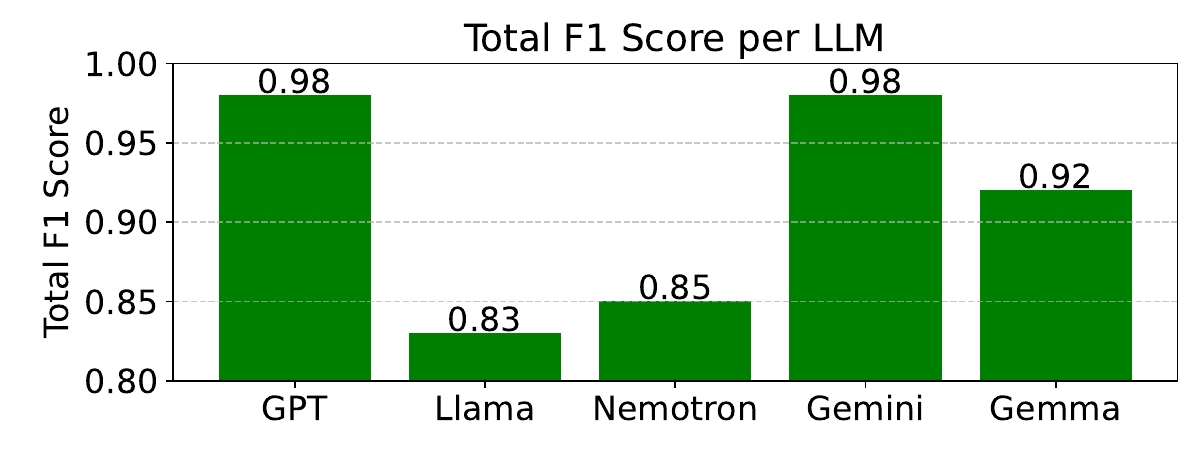}
    \caption{Comparison of the F1 Score of the LANCE implementation using GPT, Llama, Nvidia Nemotron, Gemini, and Gemma on IoC Classification.}
    \label{fig:LLMs}
\end{figure}

We evaluated the four LLMs on the BAP-generated part of the PRISM ground truth dataset. The results, summarized in Figure~\ref{fig:LLMs}, show that Gemma and Gemini perform comparably to GPT, achieving total F1 scores of $0.98$ and $0.92$, respectively. Llama and Nvidia Nemotron demonstrated moderate performance, which could likely be improved with targeted prompt engineering. A more detailed analysis of this comparison is presented in Appendix~\ref{app:Generalizability}.

In terms of coverage, Gemini and Nvidia Nemotron failed to label only 19 and 22 indicators, representing $2.1\%$ and $2.4\%$ of the dataset, highlighting their strong labeling capability relative to other leading approaches (see Section~\ref{subsec:GT_eval}). Llama and Gemma missed 84 ($9.2\%$) and 102 ($11.1\%$) indicators, respectively, indicating lower, though still substantial, labeling coverage.

In terms of processing time per report, GPT and Gemini required approximately 2–3 minutes, while Llama, Nemotron, and Gemma took around 20 minutes due to more limited computational resources. Precise timing comparisons are challenging, as external factors such as network latency and server response times can significantly influence the measurements and are not indicative of the models’ inherent performance. {The performance of Llama, Nemotron, and Gemma is expected to be substantially faster on modern cloud deployments, considering our GPUs are 5 years old. We emphasize that these latency figures are implementation artifacts, not architectural barriers, and are not indicative of fundamental scalability limitations.}

\subsection{HITL Pipeline Evaluation}
\label{LLMAndUIEval}

\subsubsection{\textbf{LANCE's Effect on Junior Analysts' Labels}}

To evaluate the LANCE-generated and the junior analysts-generated labels, we compare the Precision, Recall, and F1 scores they achieved on the 25 reports from the Baseline Annotation Pass (BAP) part of the ground truth generation methodology. We evaluate the results based on the final ground truth labels of those 25 reports.  In Figure \ref{fig:gpt_vs_analyst_perf}, we see that LANCE performed on par with the junior analysts. In the more difficult indicator types, namely domains and URLs, we see an increase of $1.9\%$ and $3.7\%$ respectively in F1 score. We also observed an increase of $1.4\%$ in total Precision, with the same metric increasing by  $5.3\%$ in domains. This means that LANCE produced fewer false positive domains than the junior analysts. Finally, we can see that LANCE achieved $6.9\%$ higher recall in URLs, producing fewer false negative URLs than the junior analysts.

\begin{tcolorbox}[boxsep=-0.5mm, boxrule=0.4mm]
\textbf{Takeaway:} LANCE-generated labels exhibit similar levels of precision, recall, and F1 score to those produced by junior analysts.
\end{tcolorbox}

In Figure \ref{fig:gpt_vs_analyst_perf}, we also see  the performance that the junior analysts and LANCE achieved on the reports from the Guided Annotation Pass (GAP) part of the ground truth creation. In this part, the junior analysts had access to the LANCE-generated labels and justifications during the labeling of the indicators. With the assistance and the additional information provided by LANCE, the junior analysts performed better in total F1 score than LANCE, and, by extension, better than without LANCE assistance. {The observed drop in domain precision in GAP is due to a higher concentration of ambiguous, legitimate domains in the GAP subset. A more detailed analysis of those indicators can be found in Section~\ref{misclassifications}.}

\begin{tcolorbox}[boxsep=-0.5mm, boxrule=0.4mm]
\textbf{Takeaway:} Junior analysts assisted by LANCE-generated labels and justifications achieve higher precision, recall, and F1 scores when labeling IoCs from unstructured threat reports. This demonstrates the practical benefit of LLM-assisted annotation workflows for cyber-threat analysts.
\end{tcolorbox}

\subsubsection{\textbf{Timing Evaluation}}

In addition to the Precision, Recall, and F1 evaluation of the analysts (junior and senior), we also evaluated the time variation due to the provided LANCE-generated labels and justification in the User Interface. We do that by automatically measuring the time each junior analyst spent on each report in BAP and GAP of the ground truth creation process. 

From the results of those measurements, we see a $43\%$ drop in the average and median time the analysts spent on a report. These findings validate a key goal of our HITL approach: significantly reducing analyst workload without compromising accuracy. The $43\%$ reduction in annotation time, when paired with consistently high precision and recall, demonstrates that LLM-augmented interfaces can streamline expert workflows by offloading routine labeling while still allowing humans to oversee critical edge cases.

\subsection{LANCE Misclassifications Analysis}
\label{misclassifications}
\begin{figure}[ht]
    \centering
    
    \fbox{\includegraphics[width=\linewidth]{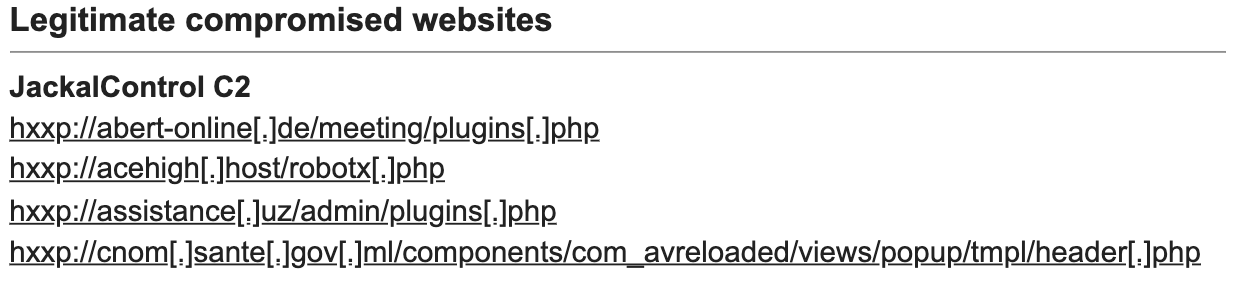}}
    
    \caption{Segment of a report included in PRISM}
    
    \label{fig:report}
\end{figure}

In this section, we perform an analysis of common misclassifications of LANCE by discussing common cases of false positives and false negatives. These cases occurred when using ChatGPT-4o~\cite{GPT4o} as the backbone LLM.
In Figure~\ref{fig:report}, we show a segment of a report which is part of PRISM dataset. Several domains and URLs of legitimate websites were compromised and used as command and control (C2) points for the campaign. According to our senior analyst, the domains (e.g., \texttt{abert-online.de}) are not IoCs (as they are a legitimate website), whereas the complete URLs are. LANCE labeled these domains as ``IoC'' and justified the label (see Table~\ref{tab:LANCEjust}) by referencing their ``\textit{involvement in the malicious activities}.'' Notably, LANCE acknowledged the domains' legitimate nature but still marked them as IoCs due to their role in the attack. The justifications for the remaining domains in Figure~\ref{fig:report} were similar, and all were consistently labeled as IoCs. In this case, both interpretations (IoC or nonIoC) can be considered correct, depending on the downstream use of the extracted indicators.
{These mislabeled cases highlight a key challenge: the definition of what constitutes an IoC can vary depending on the downstream task. For example, a compromised legitimate domain might be labeled as an IoC for detection or blocking purposes, but as nonIoC in datasets intended for long-term attribution or infrastructure tracking. To mitigate this ambiguity, systems should explicitly define IoC vs. nonIoC labels in relation to their intended use case and surface this definition to analysts during labeling. Such task-aware definitions, combined with LANCE’s explanatory justifications, can reduce false negatives by helping analysts resolve edge cases consistently and in alignment with the downstream application.}

\begin{table}[h]
    \centering
    \resizebox{\linewidth}{!}{%
    \begin{tabular}{l}
        \hline
        \textbf{LANCE justification} \\
        \hline
        The domain abert-online.de is classified as an IoC because it \\ is identified as a legitimate website that has been compromised \\ and used as a command and control (C2) server for JackalControl,\\  indicating its involvement in malicious activities.\\
        \hline
    \end{tabular}
    }
    \caption{The justification LANCE gave for the ``IoC'' label it assigned to domain \texttt{abert-online.de} which appear in the report segment shown in Figure~\ref{fig:report}.}
    \label{tab:LANCEjust}
\end{table}

{Beyond the ambiguous cases of compromised but legitimate domains, we observed two recurring categories of false negative (FN) domain labels. First, Firebase subdomains (e.g., \texttt{domain.firebase.com}) were often mislabeled as nonIoCs because the LLM defaulted to seeing Firebase as benign despite malicious context. These cases illustrate the tension between platform ubiquity and malicious abuse, and the need for downstream-task-aware definitions to guide consistent labeling.\\
Second, a systematic error emerged with domain variants containing the \texttt{www.} prefix. Indicators presented as \texttt{www.domain.com} were labeled nonIoC, while the same domain without the prefix was correctly labeled as IoC. Interestingly, in these cases, the LLM justifications often explicitly stated that the \texttt{www.} variant was malicious, even when the final label output was nonIoC. This disconnect highlights the value of explainable justifications that allowed analysts to override errors during HITL validation.}


\section{Conclusion}
\label{sec:Conclusion}

We addressed the challenge of extracting high-quality Indicators of Compromise (IoCs) from unstructured reports by introducing the first hybrid labeling system that combines explainable LLMs with human-in-the-loop (HITL) validation. Unlike naive LLM prompting or existing automated systems like VirusTotal and AlienVault, LANCE achieves high precision and recall across all IoC types while providing context-aware justifications for the analysts. Our evaluation shows that while junior analysts alone perform well, those assisted by LANCE complete tasks 43\% faster with improved accuracy, particularly on ambiguous types of indicators like domains and URLs. This human-augmented workflow enabled the creation of PRISM, the largest publicly available expert-validated IoC dataset from real-world reports. Our results show that a carefully crafted LLM-based system guided by human oversight addresses critical challenges in IoC extraction, and our dataset lays the foundation for more trustworthy and efficient methods in the future.

\bibliographystyle{IEEEtran}
\bibliography{references.bib}
\appendices

\section{LANCE Prompts}
\label{app:LANCEPrompts}

This section presents the prompt engineering phase for LANCE as well as the evaluation of the final prompts. For both the prompt engineering phase and the evaluation of the prompts, we used ChatGPT-4o \cite{GPT4o} as the underlying LLM. 

\subsection{Prompt Engineering and Input Structuring}

\begin{table*}[h]
    \centering
    \resizebox{\linewidth}{!}{%
    \begin{tabular}{lll}
        \hline
        \textbf{Prompt Technique} & \textbf{Description} & \textbf{Example} \\
        \hline
        Role specification  & Specify the role of the LLM. & You are a cybersecurity analyst.  
        
        \\
        
        Input definition  & Clear definition of what the input is.  & I will give you part of a cybersecurity report along
        \\ & & with some URLs found in that section.  
        
        \\
        
        Task definition & Clear definition of the classification task.  & Your Task is to label each of the given URLs as either ``IoC'' 
        \\ & & or ``nonIoC'' based on the context that the URL is given \\ & & in the report. 
        
        \\
        
        Definition of Terms  & Explicit definitions for ``IoC'' vs ``nonIoC''. & Label as ``IoC'' the URLs that are Indicators
        \\ &    &  of Compromise (IoCs) related to malicious activities,
        \\ & & such as phishing, malware, or other cyber threats, and as 
        \\ & & ``nonIoC'' if they are not referenced in the context of IoCs.  
        
        \\
        
        Reference to common & Descriptions of common false positives/negatives. &If a given URL is not complete (eg. is split by a new line or 
        \\
        mistakes &  & has a placeholder), do not label it as an IoC. 
        \\ & & Do not return URLs that were not in the given list, but make 
        \\ & & sure that you return all the URLs in the given list. I want you 
        \\ & & to make sure that the number of URLs in the ``Extracted URLs''
        \\ & & list that I give you is the same as the number of URLs \\ & & you return. 
        
        \\
        
        Secondary task definition & Clear definition of the justification task. & I also want you to justify your choice of the label for each URL. 
        
        \\
        
        Thoroughness Reminder & Explicit instructions to be careful and thorough. & Go through the report part twice to make sure
        \\ &  &  your labeling and justifications are correct. 
        \\ & & Make sure that for each URL you followed all the given \\ & & instructions.
        
        \\
        
        Output format request & Specification for the expected output format. & Return all the URLs, their label, and the justification
        \\ &  &  separated by ``,'', one per line, with no additional text. 
        \\ & &  The list might be empty, in that case, return an empty output.
        \\

        \hline
    \end{tabular}
    }
    \caption{LANCE prompt structure and example. The example prompt in the corresponding column is the final prompt used for LANCE for the labeling of URLs. The same techniques and structures were used as needed for the rest of the IoC Types.}
    \label{tab:Prompt}
\end{table*}

To evaluate prompt performance, we manually extracted and labeled all indicators in 10 reports. These reports were selected from AlienVault with the intent of ensuring diversity in structure, size, and IoC density. The indicator extraction and labeling process was iterative, as some indicators were missed or misclassified in the initial extraction iteration.

We conducted 27 structured prompt tests to refine our approach. Initially, we tested prompts that processed the entire report while simultaneously targeting all IoC types. However, this resulted in poor performance, including high confusion and inconsistency in model responses. This behavior has been well documented in the literature \cite{InstructionFinetuning,ZeroShot,DecomposePrompting}.To address this, we modified the prompt structure to classify each IoC type separately, leading to more reliable and consistent results. 

Each iteration involved reviewing model outputs for false positives and false negatives. Based on observed errors, we adjusted prompts for each IoC type to address recurring misclassifications. 

Due to the nature of the different types of indicators, as well as the reasons for appearing in a report, some types were more difficult to prompt than others. Specifically, domain labeling was the most difficult task to prompt-engineer. In our experiments, we came across several domain-like strings that were not actual domains. This was to be expected if taking into consideration the analysis in Section~\ref{subsec:Errors}. In addition, the appearance of nonIoC domains in IoC URLs was very common, making it necessary for the prompt to cover these instances of domains. 

The second most difficult type was URLs. The difficulty of labeling URLs arose from the fact that in several cases, there were placeholders in the extracted URLs, as they were included in the threat report as examples. Even though the obvious cases could easily be tackled with rule-based implementations, there were several cases where the example/placeholder part of the URL was not obvious or representable by a specific rule. Thus, we needed to find a way to describe the problem to the model and request it to label as ``nonIoC'' these URLs.

The prompts for IPs and file hashes were easily engineered in early iterations of the process.

During our tests, there were some occasions when we found that the prompt engineering was not enough to tackle the problem at hand, so we implemented other techniques, such as the rolling window input. We chose a size of 8000 characters, as this is close to the maximum number of tokens that state-of-the-art LLMs can output. Even though a token can include more than one character, we wanted to make sure that in cases where all 8000 characters were a list of important strings (e.g., Hashes), the model would be able to output them properly with their label. To make sure that no information would be lost due to the splitting of the report, we implemented an overlap of 50\% in each consecutive report segment \cite{LLMOverlap,promptEng1,promptEng2,promptEng3}.

The techniques analyzed in Section~\ref{subsec:LLMmethodology}, as well as the corresponding parts from the final prompt used for URLs, can be seen in Table~\ref{tab:Prompt}. By constructing our prompts this way, we address most of the pitfalls the LLM can fall into and steer it away from common inaccuracies.

We used the same set of reports to fine-tune the voting threshold for each indicator type. We initially set the voting threshold at 50\% and adjusted it iteratively to optimize performance, ensuring consistency in classification on the prompt training set.

\subsection{Prompt Evaluation}
\label{subsec:PromptEval}

\begin{table*}
\resizebox{\linewidth}{!}{%
\centering
\begin{tabular}{|l|ccccc|ccccc|ccccc|}
\hline
\textbf{Type} & \multicolumn{5}{c|}{\textbf{Precision}} & \multicolumn{5}{c|}{\textbf{Recall}} & \multicolumn{5}{c|}{\textbf{F1-Score}} \\
 & \textbf{GPT} & \textbf{Llama} & \textbf{Nemotron} & \textbf{Gemini} & \textbf{Gemma} & \textbf{GPT} & \textbf{Llama} & \textbf{Nemotron} & \textbf{Gemini} & \textbf{Gemma} & \textbf{GPT} & \textbf{Llama} & \textbf{Nemotron} & \textbf{Gemini} & \textbf{Gemma} \\
\hline
URL    & 0.97 & 0.71 & 0.80 & 0.97 & 0.90 & 0.99 & 0.35 & 0.52 & 0.98 & 0.84 & 0.98 & 0.47 & 0.63 & 0.97 & 0.87 \\
Domain & 0.96 & 0.83 & 0.76 & 0.94 & 0.75 & 0.94 & 0.78 & 0.78 & 1.00 & 0.77 & 0.95 & 0.80 & 0.77 & 0.97 & 0.76 \\
HASH   & 1.00 & 1.00 & 1.00 & 1.00 & 1.00 & 1.00 & 0.86 & 1.00 & 0.99 & 0.99 & 1.00 & 0.93 & 1.00 & 0.99 & 0.99 \\
IP     & 1.00 & 1.00 & 1.00 & 1.00 & 0.99 & 0.95 & 0.99 & 0.68 & 0.98 & 0.95 & 0.97 & 0.99 & 0.81 & 0.99 & 0.97 \\
\hline
\textbf{Total} & \textbf{{0.98}} & \textbf{0.92} & \textbf{0.90} & \textbf{{0.98}} & \textbf{0.93} & \textbf{0.98} & \textbf{0.76} & \textbf{0.80} & \textbf{{0.99}} & \textbf{0.91} & \textbf{{0.98}} & \textbf{0.83} & \textbf{0.85} & \textbf{{0.98}} & \textbf{0.92} \\
\hline
\end{tabular}
}
\caption{Comparison of the LANCE implementation using GPT, Llama, Nvidia Nemotron, Gemini, and Gemma on IoC Classification.}
\label{tab:LLM_eval}
\end{table*}

For the evaluation of the automated IoC extraction pipeline, we collected the 10 most recent reports added to the AlienVault database from their original user profile. We then manually extracted and labeled the indicators that existed in the text of the reports to create our test dataset.

\begin{figure}[ht]
    \centering
    
    \includegraphics[width=\linewidth]{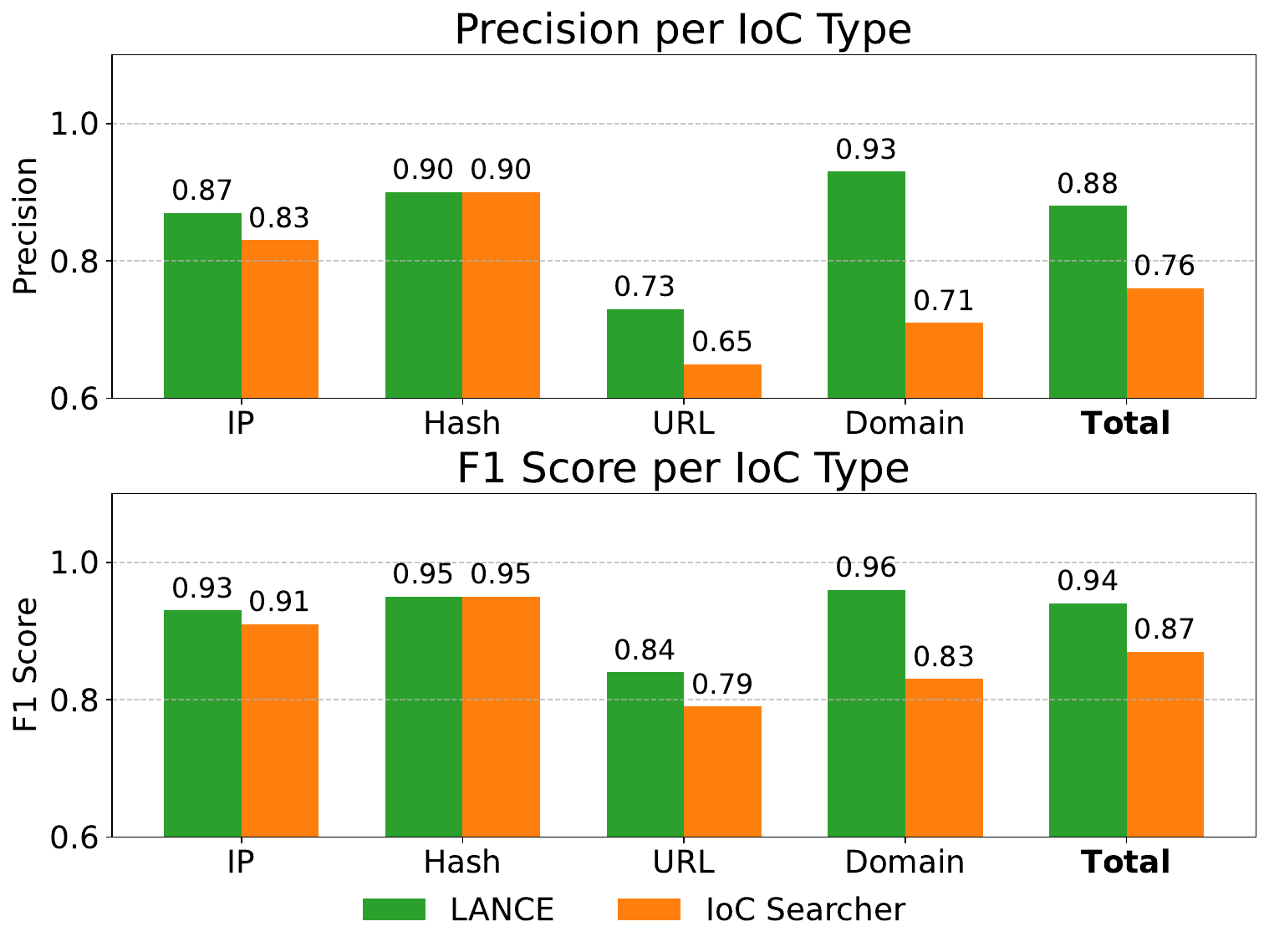}
    
    \caption{Comparison of precision and F1 score of LANCE and IoC Searcher across multiple IoC types.}
    
    \label{fig:promt_eval}
\end{figure}

We evaluated the Precision, Recall, and F1 scores of LANCE and compared them with IoC Searcher. The evaluation can be seen in Figure \ref{fig:promt_eval}. The Recall of both LANCE and IoC Searcher was 100\% in all types of IoCs and in total. LANCE consistently achieves higher precision and F1, particularly on domains and URLs. The values above each bar indicate the exact performance for each indicator type and overall.

\section{Evaluation of IoC extraction components of CTI extraction tools}
\label{app:CTI_eval}

{Table~\ref{tab:CTI_eval} presents the precision, recall, and F1 scores of the IoC extraction components of representative prior systems that focus on CTI extraction on PRISM. Extractor~\cite{extractor}, LADDER~\cite{LADER}, and CTINexus~\cite{CTINexus} were originally designed for broader CTI extraction tasks such as attack pattern and knowledge graph construction. Their indicator extraction components rely primarily on regular expressions and are therefore not optimized for precise IoC labeling. This limitation is evident in their relatively low F1 scores on PRISM, with CTINexus in particular achieving very poor recall. IoCSearcher~\cite{GoodFATR}, a state-of-the-art RegEx tool, attains perfect recall but at the cost of precision. In contrast, LANCE balances both dimensions, achieving an F1 score of 0.98. These results underscore that while prior CTI systems contribute valuable higher-level intelligence, reliable IoC extraction and ground truth creation require special attention. Moreover, integrating LANCE into such systems could substantially improve their IoC extraction performance, providing them with a more accurate foundation on which to build higher-level CTI analysis.}

\begin{table}[ht]
\centering
\begin{tabular}{|l|c|c|c|}
\hline
\textbf{Tool} & \textbf{Precision} & \textbf{Recall} & \textbf{F1} \\
\hline
IoCSearcher \cite{GoodFATR} & 0.44 & \textbf{1.00} & 0.61 \\
\hline
Extractor \cite{extractor} & 0.38 & 0.32 & 0.35 \\
LADDER \cite{LADER} & 0.70 & 0.37 & 0.48 \\
CTINexus REGEX only \cite{CTINexus} & 0.30 & 0.33 & 0.31 \\
CTINexus \cite{CTINexus} & 0.52 & 0.02 & 0.03 \\
\hline
LANCE & \textbf{0.98} & 0.98 & \textbf{0.98} \\
\hline
\end{tabular}
\caption{Precision, Recall, and F1 score for IoC extraction components of CTI extraction tools on PRISM.}

\label{tab:CTI_eval}
\end{table}
\section{LANCE Generalizability}
\label{app:Generalizability}


In Table~\ref{tab:LLM_eval} we see a detailed evaluation of LANCE using 5 state of the art LLMs for the LLM component: ChatGPT 4o \cite{GPT4o}, LLama 3.3 70b \cite{LLama3.3}, Nvidia Llama 3.1 Nemotron 70b \cite{nvidiallama}, Gemini 2.0 Flash \cite{gemini2flash}, and Gemma 3 27b \cite{gemma3}. The table shows the precision, recall, and F1 score the pipeline achieved with each model per indicator type and in total. The total F1 score for each LLM is also shown in Figure~\ref{fig:LLMs}.

\section{Ground truth creation methods evaluation details}
\label{app:GTevalDetails}

In Table~\ref{tab:GT_creation_method_eval_on_GT} we see the performance of the five ground truth creation methods discussed in Section~\ref{subsec:GT_eval} on coverage as well as Precision, Recall, and F1 score on each type of IoCs (IPs, hashes, URLs, and domains).

\begin{table*}[ht!]
    \resizebox{\linewidth}{!}{%
    \centering
    \begin{tabular}{|ll|cc|ccc|ccc|ccc|ccc|}
        \hline
        \multirow{2}{*}{Method} && Labeled & Unlabeled  & \multicolumn{3}{c}{IP} & \multicolumn{3}{c}{Hash} & \multicolumn{3}{c}{URL} & \multicolumn{3}{c|}{Domain} \\
        \cline{5-16}
        &&Indicators&Indicators& Precision & Recall & F1 & Precision & Recall & F1 & Precision & Recall & F1 & Precision & Recall & F1 \\
        \hline
        
        \multicolumn{2}{|l|}{RegEx + Whitelists \cite{GoodFATR}}  & 1789 & 0 & 0.987 & \textbf{{1.000}} & \textbf{{0.994}} & 0.994 & \textbf{{1.000}} & 0.997 & 0.720 & \textbf{{0.992}} & 0.835 & 0.523 & \textbf{{1.000}} & 0.687 \\
        \hline

        \multicolumn{2}{|l|}{AlienVault \cite{iACE}}  & 1464 & {653} & 0.936 & 0.948 & 0.942 & 0.686 & 0.965 & 0.802 & 0.855 & 0.256 & 0.394 & 0.928 & 0.737 & 0.821 \\
        \hline

        \multicolumn{1}{|l!{\vrule}}{RegEx+VT}  &th=1 \cite{Twiti, IoCStalker} & 1502 & 287 & 0.993 & 0.929 & 0.960 & \textbf{{0.998}} & 0.674 & 0.805 & 0.965 & 0.863 & 0.911 & 0.826 & 0.961 & 0.888 \\

        \multicolumn{1}{|l!{\vrule}}{\cite{AspIoC, HinCTI, Extracting_actionable_information}}  &th=5 \cite{IoCStalker} & 1502 & 287 & \textbf{{1.000}} & 0.503 & 0.670 & \textbf{{0.998}} & 0.626 & 0.769 & \textbf{{0.995}} & 0.785 & 0.878 & \textbf{{0.891}} & 0.730 & 0.802 \\

        \hline
        \multicolumn{2}{|l|}{\textbf{LANCE}} & 1774 & {15} & \textbf{{1.000}} & 0.968 & 0.984 & 0.996 & \textbf{{1.000}} & \textbf{{0.998}} & 0.970 & \textbf{{0.992}} & \textbf{{0.981}} & 0.878 & 0.950 & \textbf{{0.913}} \\
        \hline
    \end{tabular}
    }

    \caption{Performance comparison of four prominent ground truth creation methods from Table~\ref{tab:ioc_tools}, across four IoC types (IP, hash, URL, domain). The RegEx + Whitelists methodology \cite{GoodFATR} exhibits near-perfect recall but low precision due to excessive false positives. VirusTotal-based methods vary depending on the chosen threshold, but they have a high number of unlabeled indicators. The same can be said for AlienVault, which has the highest number of unlabeled indicators. LANCE achieves the highest F1 score across all indicator types while providing excellent coverage.}
    \label{tab:GT_creation_method_eval_on_GT}
\end{table*}

\section{User Interface}
\label{app:UI}

\begin{figure*}
    \centering
    \includegraphics[width=0.8\textwidth]{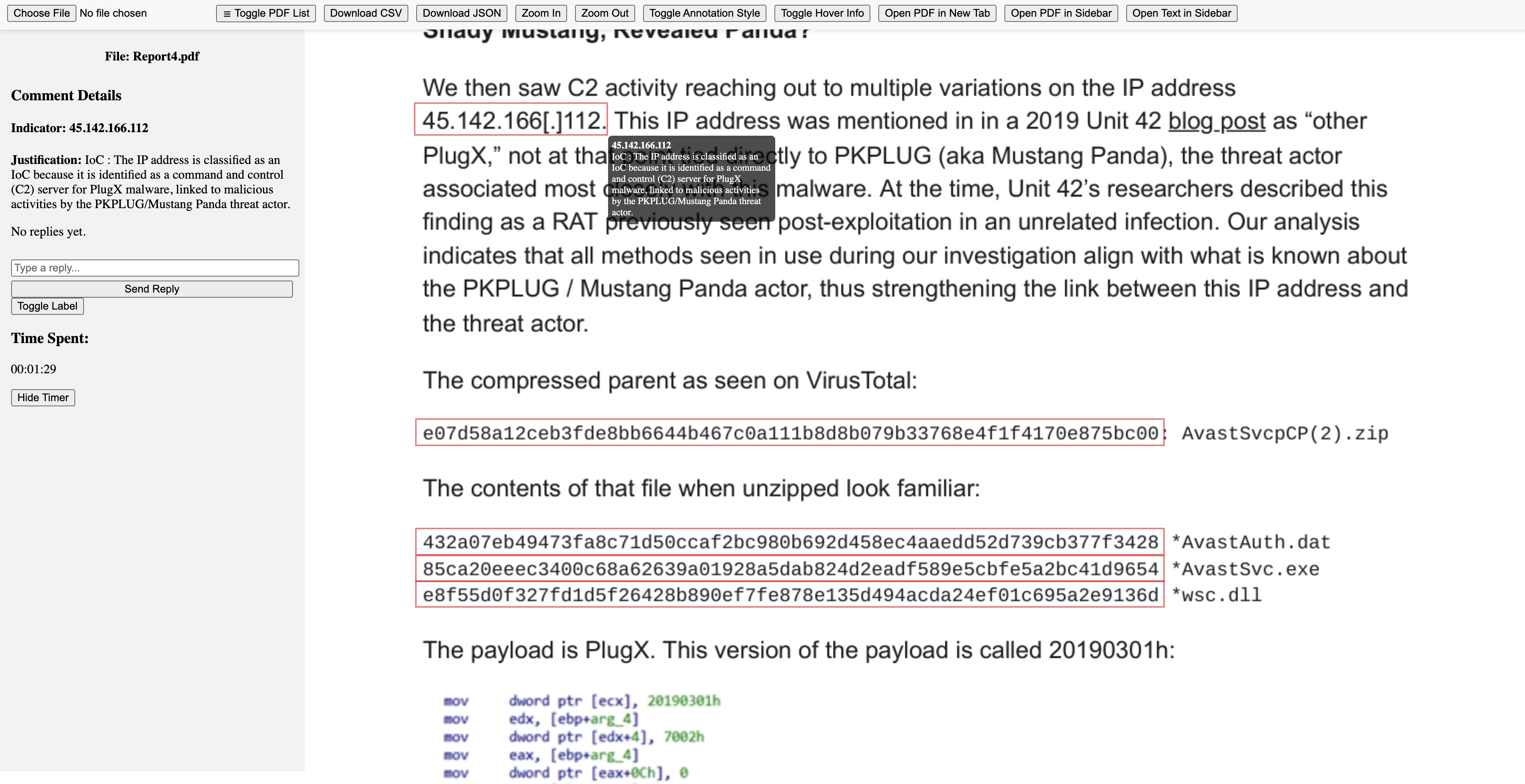}
    \caption{This is a view of the user interface that the analysts used for the GAP of the ground truth creation. Contrary to the User Interface for BAP, the indicators are extracted and labeled by LANCE. The indicators labeled as IoC
    can be seen in a red rectangle, while those labeled as nonIoC can be seen in a green rectangle. A justification is provided for all indicators.
    }
    \label{fig:UI2}
\end{figure*}

\begin{figure*}
    \centering
    \includegraphics[width=0.8\textwidth]{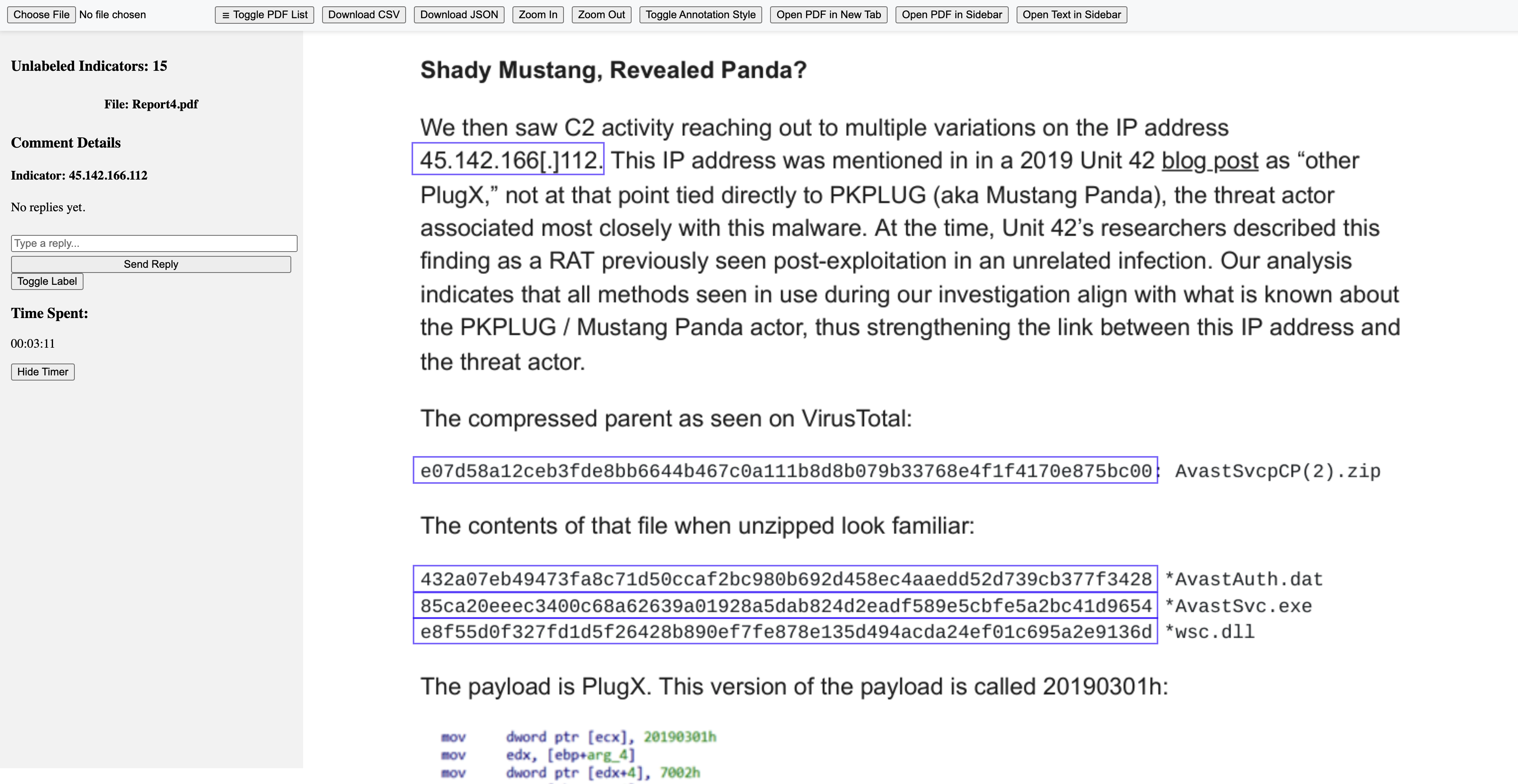}
    \caption{This is a view of the user interface that the analysts used for the BAP of the ground truth creation. The extracted Indicators can be seen marked in a blue rectangle, which changes color according to the label assigned by the analyst.}
    \label{fig:UI1}
\end{figure*}

Figures \ref{fig:UI1} and \ref{fig:UI2} show a view of the custom user interface developed for our system. In the proposed user interface (Figure~\ref{fig:UI2}), the analyst can see all indicators highlighted by a red or green rectangle. In the cases that the LANCE-generated label is \textit{IoC}, the rectangle is red, while if the label is \textit{nonIoC} it is green \cite{UI1,UI2,UI3}. This way, the analyst can quickly understand the generated label. By hovering over the indicator or by selecting the indicator, they can see the justification that LANCE provided for the selected label.  The label of each indicator can be changed by the analyst either by selecting the indicator and clicking the designated button or by double-clicking the indicator. The label change propagates to all instances of this indicator throughout the report. 

For the baseline annotation pass (BAP), the user interface was altered to not show the designated colors unless the label was assigned by the analyst. The analyst could also not see the label, justification, or any other LANCE-generated information about any indicator. For this pass, we also implemented a counter that showed the remaining unlabeled indicators.

\end{document}